\documentclass[a4paper,fleqn,usenatbib]{mnras}

\usepackage{newtxtext,newtxmath}

\usepackage[T1]{fontenc}
\usepackage{ae,aecompl}

\usepackage{graphicx}	
\usepackage{amsmath}	
\usepackage{amssymb}	
\usepackage{hyperref}
\usepackage{epstopdf}
\usepackage{bm}

\newcommand{\Kepler}{\textit{Kepler~}}
\newcommand{\Rho}{\mathrm{P}}

\title[Extracting Information from AGN Variability]{Extracting Information from AGN Variability}

\author[V.~P. Kasliwal et al.]{
\parbox{\textwidth}{
Vishal~P. Kasliwal,$^{1,2}$\thanks{E-mail: \texttt{vish@sas.upenn.edu}}
Michael~S. Vogeley,$^{3}$
\& Gordon~T. Richards$^{3}$}\vspace{0.4cm}\
\\
\parbox{\textwidth}{
$^{1}$Department of Physics \& Astronomy, University of Pennsylvania, 209 South 33rd Street, Philadelphia, PA 19104-6396, USA\\
$^{2}$Department of Astrophysical Sciences, 4 Ivy Lane, Princeton University, Princeton, NJ, 08544, USA\\
$^{3}$Department of Physics, Drexel University, Philadelphia, PA 19104-2875, USA\\
}
}

\date{Accepted XXX. Received YYY; in original form ZZZ}

\pubyear{2017}

\begin{document}
\label{firstpage}
\pagerange{\pageref{firstpage}--\pageref{lastpage}}
\maketitle

\begin{abstract}
AGN exhibit rapid, high amplitude stochastic flux variations across the entire electromagnetic spectrum on timescales ranging from hours to years. The cause of this variability is poorly understood. We present a Green's Function-based method for using variability to (1) measure the time-scales on which flux perturbations evolve and (2) characterize the driving flux perturbations. We model the observed light curve of an AGN as a linear differential equation driven by stochastic impulses. We analyze the light curve of the \Kepler AGN Zw 229-15 and find that the observed variability behavior can be modeled as a damped harmonic oscillator perturbed by a colored noise process. The model powerspectrum turns over on time-scale $385$~d. On shorter time-scales, the log-powerspectrum slope varies between $2$ and $4$, explaining the behavior noted by previous studies. We recover and identify both the $5.6$~d and $67$~d timescales reported by previous work using the Green's Function of the C-ARMA equation rather than by directly fitting the powerspectrum of the light curve. These are the timescales on which flux perturbations grow, and on which flux perturbations decay back to the steady-state flux level respectively. We make the software package \href{https://github.com/AstroVPK/kali}{\textsc{k\={a}l\={i}}} used to study light curves using our method available to the community.
\end{abstract}




\section[Introduction]{Introduction}\label{sec:Introduction}

Active Galactic Nuclei (AGN) exhibit variability on time-scales ranging from minutes and hours to years over the full electromagnetic spectrum. The underlying cause of the variability is not clear. Proposed models range from local viscosity fluctuations to oscillatory modes in the accretion disk \citep{UrryARAA}. AGN are powered by the accretion of matter onto a central supermassive black-hole \citep{ReesARAA}. Matter infalling onto the supermassive black hole must lose angular momentum during inspiral. The loss of angular momentum results in the conversion of gravitational binding energy to the kinetic energy of the flow. A portion of this kinetic energy is radiated away resulting in the characteristic appearance of the AGN: high luminosity ($\sim 10^{47}$~erg~s$^{-1}$ for the brightest quasars) emanating from a small volume (radius $r \sim 10^{14}$~cm) of space \citep{Edelson96}.

It is generally accepted that the accretion inflow results in the formation of an accretion disk \citep*{KoratkarBlaes99,PringleARAA}. At least three mechanisms may be responsible for the extraction of angular momentum from the accretion flow in a standard disk: (1) Magneto-rotational Instability (MRI) generated turbulence \citep*{BHI,BHRev}; (2) Large scale magneto-hydrodynamic (MHD) outflow caused external stresses \citep*{BlandfordPayne82}; and (3) Shocks produced by non-axis-symmetric waves \citep{FragileBlaes08}. MRI prescriptions for angular momentum transport are succinctly quantified by the alpha-prescription of \citet{ShakuraSunyaev73}---see \citet{BalbusPapaloizou99}. Under the Shakura-Sunyaev $\alpha$-prescription, accretion flows are modelled as solutions of the vertically integrated hydrodynamic equations for the conservation of mass, radial and angular momentum, and energy. Stability criteria then dictate that solutions be of one of three types as characterized by location in surface density-mass accretion rate phase space \citep{BlaesAccretion}. Low accretion rate coupled with low surface density results in the formation of advection-dominated accretion flows \citep{NarayanYi94,ChenAbramowicz95} where the cooling time of the accretion flow is much greater than the in-fall time. Such disks are optically thin and may exist in low-luminosity AGN. Some quasars, a few Seyfert 1s and the intermediate states of black-hole X-ray binaries (BHXRB) are thought to possess radiation pressure dominated `slim' disks---$H/r < 1$ where $H$ is the local disk height at radius $r$---with medium surface mass densities and very high accretion rates \citep{Abramowicz88}. Such disks also have shorter inflow time as compared to the cooling time-scale, making the flow advective. The third stable solution occurs at high surface mass density but low accretion rate and results in the formation of classic `thin' accretion disks that are optically thick and geometrically thin ($H/r << 1$) \citep*{ShakuraSunyaev73,FKR}. Most AGN and BHXRBs in the high and soft states are thought to possess such disks \citep{BlaesAccretion}.

Numerous mechanisms have been posited for producing variability in accretion disks---see \citet{DoneAccretion}, \citet{MaccaroneAccretion}, and \citet{UttleyAccretion} for comprehensive overviews. While we are concerned primarily with accretion onto supermassive black holes (SMBH) in AGN, a fair fraction of the theory is developed with BHXRB accretion in mind. Although similarities exist between accretion in BHXRBs and AGN, the difference in energy densities is considerable, with AGNs having considerably higher luminosities and exhibiting variability on longer time-scales. For instance, quasi-periodic oscillations (QPO) are routinely observed in BHXRBs, whereas in AGN there are few reports of QPOs \citep{GierlinskiQPO,Andrae13,GrindleyQPO}. While BHXRBs are commonly observed to shift between various spectral states, such transitions have not been observed in AGN to date \citep{Kelly11}.

AGN optical variability may arise due to fluctuations in the \citet{ShakuraSunyaev73} viscosity parameter $\alpha$. \citet{Lyubarskii97} showed that local stochastic fluctuations in the accretion disk are capable of producing the power-law power spectral densities (PSD) observed in AGN and BHXRBs. Such fluctuations are thought to propagate inwards over time, suggesting that a powerful observational test is the observation of blue-lags \citep{UttleyAccretion}, i.e., light curve variations at shorter wavelengths lagging those at longer wavelengths. Evidence for the propagating fluctuation model was provided by \citet{Miyamoto88} who observed that hard X-ray variations in Cyg X-1 lagged behind soft X-ray variations with an inverse scaling between the time lag and the variability time-scale. \citet{Starling04} used the viscosity fluctuation model to impose lower limit of $\alpha \sim 0.01$ on the disk viscosity. A later study by \citet{King07} using AGN variability triggered by viscosity fluctuations suggests that $\alpha \sim 0.1$-$0.4$. \citet{Wood01} examined a variant of the \citet{Lyubarskii97} propagating fluctuations model in which episodic mass deposition at the outer edge of the accretion disk triggers viscosity fluctuations in the accretion disk. This model was revisited by \citet{Titarchuk07} who generalized the treatment of \citet{Wood01} to include time-dependent fluctuation sources located randomly on the disk face. \citet{Titarchuk07} follow the evolution of perturbations to the mass surface density and luminosity arising from viscosity that is a power-law function of disk radius. Subsequently, \citet{Kelly11} studied solutions to the surface mass density equation perturbed by stochastic fluctuations and showed that the resulting light curve can be expressed as a linear combination of multiple Ornstein-Uhlenbeck processes \citep{Gillespie96,Kelly09}.

Accretion disk variability may also be caused by the existence of `hot-spots' in the accretion disk (e.g., \citealt{MaccaroneAccretion}). Hot spots are thought to arise when accretion streams impact the disk resulting in shock heating. Such hot spots have already been detected during periods of quiescence in BHXRBs \citep{Froning11,McClintock95}.

Thermal viscous instabilities or a variable mass transfer rate may be responsible for accretion disk variability \citep{Lasota01,Coriat12}, though the amplitude of the variability should be much smaller and the time-scale much longer for AGN than for BHXRBs \citep{Hameury09}.

Other sources of variability include magnetic flares in the corona of the accretion disk \citep{PoutanenFabian99} or from the hot accretion flow itself \citep{Veledina13}. Alternatively, dynamo processes due to the magnetic fields threading the disk may be responsible for producing variability \citep{LivioPringleKing03,King04,MayerPringle06}. \citet{JaniukCzerny07} have suggested that the variability may be produced in the hot X-ray corona surrounding the accretion disk from the evaporation of mass perturbations in the disk. \citet{Misra08} suggested that the X-ray variability of Cyg X-1 can be modeled as a damped harmonic oscillator perturbed by noise. Quasi-periodic signals in the light curve may be produced by weak shocks in the innermost region of the accretion caused by the presence of a tilt in the accretion disk of the black hole \citep{FragileBlaes07,FragileBlaes08}.

Quasi-periodic signals may be caused by global oscillation modes or parametric resonance instability \citep{ReynoldsMiller09a,ReynoldsMiller09b,Oneill11}. Variability may also be produced by standing shocks interacting with fluctuations in tilted accretion disks \citep{Henisey12}.

Numerical simulations of light curves from accretion disks began appearing in the literature starting with \citet{Schnittman06}. A ray-tracing and radiative transfer code was used to create images of a simulated accretion disk. Using images generated during multiple time dumps in the simulation, \citet{Schnittman06} were able to create mock light curves resulting from accretion flows. \citet{Noble09} performed fully General-Relativistic (GR) magneto-hydrodynamic (MHD) simulations of accretion flows around black holes and showed that while the variability was generated by accretion rate variances caused by MHD turbulence, time delay effects steepened the PSD of the variations. Other GRMHD simulations have also shown that MHD turbulence results in accretion rate variations that can produce variability \citep{Moscibrodzka09,Dexter09,Dexter10}. Recently, \citet{Schnittman13a} and \citet{Schnittman13b} have even managed to produce full (X-ray) spectra demonstrating the effect of variability on the spectral slope.

Several mechanisms may be responsible for the variability observed in AGN light curves. Light curves in the optical are thought to be driven by X-ray variability on shorter time-scales but exhibit a local source of variability on longer time-scales \citep{UttleyAccretion}. This implies that it is unlikely that a single mechanism drives variability across the entire electromagnetic spectrum. Rather, the nature of the underlying mechanism responsible for variability may be different at different wavelengths.

Studies of AGN variability model the observed light curve as stochastic processes i.e mathematically the light curve as a function of time is a random variable that correlates with previous values of the same light curve. Prescriptions for stochastic models require two properties to be specified: (1) the distribution from which to draw the random values of the light curve; and (2) the strength of the correlation of a given value of the light curve with previous values of the same light curve. Most studies of AGN variability (implicitly) invoke some form of the central limit theorem to argue that property 1 should be a Gaussian distribution which then lets them specify property 2 using a second-order statistic such as the power spectral density (PSD), or the auto-covariance function (ACVF), or the structure function (SF). The most straightforward method of fitting such a stochastic model to a light curve is by direct estimation of the PSD of the observed light curve. Using periodogram estimators of the PSD, the observed PSD can be compared to the predicted theoretical PSD as advocated in \citet*{Uttley02}. Alternatively, it is also possible to estimate the ACVF of the observed light curve and fit it to the theoretical ACVF of a given stochastic process. Equivalently, following \citet*{Kasliwal15}, estimates of the structure function (SF) of the observed light curve can be fit to the theoretical SF computed either directly from the theoretical ACVF, or from Monte-Carlo simulations of light curves generated using the prescribed model PSD.

The principle drawback with the methods above is that while stochastic models provide expressions for the analytic forms of the PSD, ACF, and SF, typically no analytic results exist about the \textit{distributions} of the estimates of these statistics about the theoretical curve \citep{BrockwellDavisITSF}. Even simple assumptions about the Gaussianity of the distribution of the estimates may be incorrect. For example, as shown in \citet*{Emm10} and in \citet{Kasliwal15}, it is the estimates of the logarithm of the SF rather than the SF that are Gaussian distributed. This means that the application of the methods above rely on Monte-Carlo techniques to estimate the distribution of the property, be it the PSD, ACVF or SF that is being fit. Usually, some assumption of normality is made and typically the estimates are treated as independent purely for computational practicality.



Recently, \citet{Kelly14} suggested that stochastic variability in a large class of astrophysical sources may be modelled using Continuous-time AutoRegressive Moving Average (C-ARMA) processes. The attractiveness of using C-ARMA processes to model stochastic phenomenon stems from two factors: (1) C-ARMA processes directly specify property 2, i.e., the correlation structure of the light curve by providing a differential equation that governs how the light curve evolves over time. This alternative to providing a second-order statistic is powerful because it makes the light curve evolution more amenable to physical interpretation as is shown in this work. It also provides some idea of how to replace Gaussian distributions with more generalized $\alpha$-stable distributions that may be truer to reality \citep{BrockwellMarquardt05, BrockwellLinder09}; and (2) flexibility of the shape of their PSD. Since the PSD of a C-ARMA process is a rational function, it can be made arbitrarily complex to model a very wide range of phenomena. This flexibility is achieved by allowing C-ARMA processes to have multiple Autoregressive (quantified by $p$) and Moving Average (quantified by $q$) terms under the constraint $q < p$. If $\chi$ is a small flux perturbation from the mean flux level, then a C-ARMA($p$,$q$) process that models the time evolution of $\chi$ has the general form
\begin{multline}\label{eq:CARMAIntro}
\mathrm{d}^{p}\chi + \alpha_{1} \mathrm{d}^{p-1}\chi + \ldots + \alpha_{p-1} \mathrm{d}\chi + \alpha_{p} \chi = \\ \beta_{0} (\mathrm{d}W) + \beta_{1} \mathrm{d}(\mathrm{d}W) + \ldots + \beta_{p-1} \mathrm{d}^{p-1}(\mathrm{d}W)
\end{multline}
where $W$ is a \textit{Wiener process} and is responsible for introducing stochasticity into the behavior of $\chi$. The $\alpha$ and $\beta$ are constants that govern the behavior of the resulting process and have dimensions of various powers of time. Several studies have found that C-ARMA processes model AGN variability well \citep{Simm16, Caplar17}. In this paper, we present a Green's function-based method for understanding what a C-ARMA process tells us about accretion physics.

\section[Outline]{Outline}\label{sec:Outline}

We propose a Green's function-based method for analysing observed light curves, particularly in the UV through IR where the fluctuations are smaller than they tend to be in the X-ray. Such a method facilitates comparison with analytic and simulation work on variability. Our method builds on the C-ARMA modelling technique of \citet{Kelly14} by demonstrating how the Green's Function of the C-ARMA process left hand side (LHS) is a powerful probe of the physics driving the variability.

We discuss the Green's Function approach to studying AGN variability as modelled by C-ARMA processes in section~\ref{sec:Meaning} and present the underlying mathematical details in appendices~\ref{sec:LHS},~\ref{sec:GFComputation}, and~\ref{sec:RHS}. A simple example of a C-ARMA process---the C-ARMA($2$,$1$) process---is presented in section~\ref{sec:CARMA21}. We have written a fast but easy-to-use software package, \href{https://github.com/AstroVPK/kali}{\textsc{k\={a}l\={i}}}, for inferencing C-ARMA model parameters from observed light curves. We present an overview of the operation of \href{https://github.com/AstroVPK/kali}{\textsc{k\={a}l\={i}}} in section~\ref{sec:Steps} and defer mathematical details to appendix~\ref{sec:kali}.

Appendices~\ref{sec:Fitting} through~\ref{sec:ModelSelection} present details of inferring the properties of the C-ARMA process from noisy observations using the Kalman filter. Alternatives exist to using the Kalman filter for inferring the parameters of a C-ARMA process. We discuss why the Kalman filter is the optimal choice of fitting techniques in appendix~\ref{sec:Fitting}
and suggest an alternative to the traditional \textit{controllable canonical form} state-space representation used by \citet{JonesAckerson90} and \citet{Brockwell14}. Our choice of state-space representations is (1) better suited for extension to more complex models in the future and (2) computationally cheaper for regularly sampled data. The Kalman filter is presented in appendix~\ref{sec:Kalman}. In appendix~\ref{sec:ModelSelection}, we discuss information criteria that can be used to select the model order.

We present an example of the application of C-ARMA modelling to the \Kepler light curve of the Seyfert 1 galaxy Zw 229-15 in section~\ref{sec:Zw229-15}. In this section, we demonstrate how the C-ARMA model can be used to explain the multiple slopes and time-scales found in the light curve of this AGN by previous work. We present our conclusions in section~\ref{sec:Conclusions}.

\section[C-ARMA Physics]{Physics from the C-ARMA Process Equation}\label{sec:Meaning}


We present motivations for using C-ARMA processes to model AGN variability. While the physical origins of AGN variability remain unclear, the phenomenon is thought to arise in the accretion disk of the AGN. Radiation from the accretion disk is further modulated by matter in the region surrounding the disk, forming a complex system. The flux, $F(t)$, emitted by the system as a whole is known to vary non-linearly, with stronger variability being observed when the object is brighter \citep{UMV05}. Assume that there exists some stochastic differential equation that phenomenologically models $F(t)$ over the duration that the AGN system is observed for. The non-linearity observed in AGN variability suggests that any phenomenological model of variability be non-linear. However if the variations in the flux are small, such as are observed in the optical, variability can be modeled as perturbations; we may write $F(t) = F_{0} + \chi(t)$ where the $\chi(t)$ are perturbations to the steady-state flux $F_{0}$. We can then \textit{linearise} the non-linear equation that governs $F(t)$ and obtain a \textit{linear} equation for the perturbations $\chi(t)$. In appendix~\ref{sec:LHS}, we show how the linearisation of an arbitrary non-linear equation for the flux $F(t)$ yields the left hand side (LHS) of the C-ARMA equation,
\begin{equation}\label{eq:LHS}
\mathrm{d}^{p}\chi + \alpha_{1} \mathrm{d}^{p-1}\chi + \ldots + \alpha_{p-1} \mathrm{d}\chi + \alpha_{p} \chi = u.
\end{equation}
Here, $u$ are input impulses that drive the behavior of the flux perturbations, $\chi$.

The impulse response of the system is characterized by the Green's function of equation~\eqref{eq:LHS}. Particular solutions are found by linear superposition (integration) of the Green's function over all the impulses that drive the system \citep{PanditWu}. Here, the Green's function quantifies how a unit flux perturbation evolves after it is triggered. By comparing the observed Green's function to the expected behaviour of perturbations from theoretical considerations, we can gain insight into accretion physics. The form of the Green's function for equation~\eqref{eq:LHS} is given by
\begin{equation}\label{eq:GFSol}
G(t) = \sum_{k = 1}^{p} c_{k} \mathrm{e}^{\rho_{k}t},
\end{equation}
where the $\rho_{k}$ are the roots of the autoregressive polynomial of equation~\eqref{eq:ARCharPoly} and the $c_{k}$ are constants that can be determined as stated in appendix~\ref{sec:GFComputation}. Using Euler's formula, it is clear that the real part of each $\rho_{k}$ corresponds to an exponential decay timescale $\tau_{k} = 1/\Re(\rho_{k})$ while the complex part of each $\rho_{k}$ produces modulating oscillations with period $T_{k} = 2\upi/\Im(\rho_{k})$.

The cause of perturbations, i.e. variability, in the flux emanating from an AGN is unknown; the C-ARMA model of \citep{Kelly14} chooses a very general mathematical model for the form of the flux perturbations, 
a correlated Gaussian noise process given by
\begin{equation}\label{eq:RHS}
u = \beta_{0} (\mathrm{d}W) + \beta_{1} \mathrm{d}(\mathrm{d}W) + \ldots + \beta_{q-1} \mathrm{d}^{q-1}(\mathrm{d}W) + \beta_{q} \mathrm{d}^{q}(\mathrm{d}W),
\end{equation}
with $q < p$. In appendix~\ref{sec:RHS}, we present an overview of the mathematics of Weiner processes and increments and discuss why it is necessary to impose the condition that $q < p$. Alternatives exist to the C-ARMA process choice of driving impulses such as models in which the magnitude of the driving impulses is determined by some function of the current value of the flux such as in the Cox-Ingersoll-Ross model \citep{CoxIngersollRoss85}.

The form for the input stochastic impulses can be characterized by the power spectral density (PSD) of the impulses given by
\begin{equation}\label{eq:ImpulsePSD}
S_{uu}(\nu) = \frac{1}{2 \upi } |\beta(2 \upi \imath \nu)|^{2},
\end{equation}
where $\beta(z)$ is the moving average polynomial of equation~\eqref{eq:MACharPoly}. The PSD of the stochastic impulses tells about the timescales on which the stochastic impulses occur. Combining equation~\eqref{eq:LHS} with equation~\eqref{eq:RHS} yields the original C-ARMA process equation~\eqref{eq:CARMAIntro}. A C-ARMA process is a linear differential equation perturbed/driven by a random process. \textit{We see that C-ARMA processes attempt to describe AGN luminosity variations in terms of a linear system driven by colored noise}.

Physically, the impulses could potentially be attributable to local `hot-spots' in the accretion disk---in which case, the linear differential equation describes how the hot spots evolve and dissipate. In this picture, the Green's function of the linear differential equation is the impulse-response function of the accretion disk material to small perturbations of the luminous flux. The driving process may be viewed as disturbances generated by MRI instabilities in the accretion disk. If accretion disk hot-spots are the root cause of AGN variability, the powerspectrum of the driving stochastic impulses characterizes the timescales on which the hot-spots evolve.


In the next section we discuss a simple C-ARMA process---the C-ARMA($2$,$1$) process---in detail and derive the characteristic time-scales present in this process as an example.

\section[C-ARMA($2$,$1$)]{An Example of a C-ARMA Process: The C-ARMA($2$,$1$) Process}\label{sec:CARMA21}

\begin{figure*}
    \includegraphics[width=\textwidth]{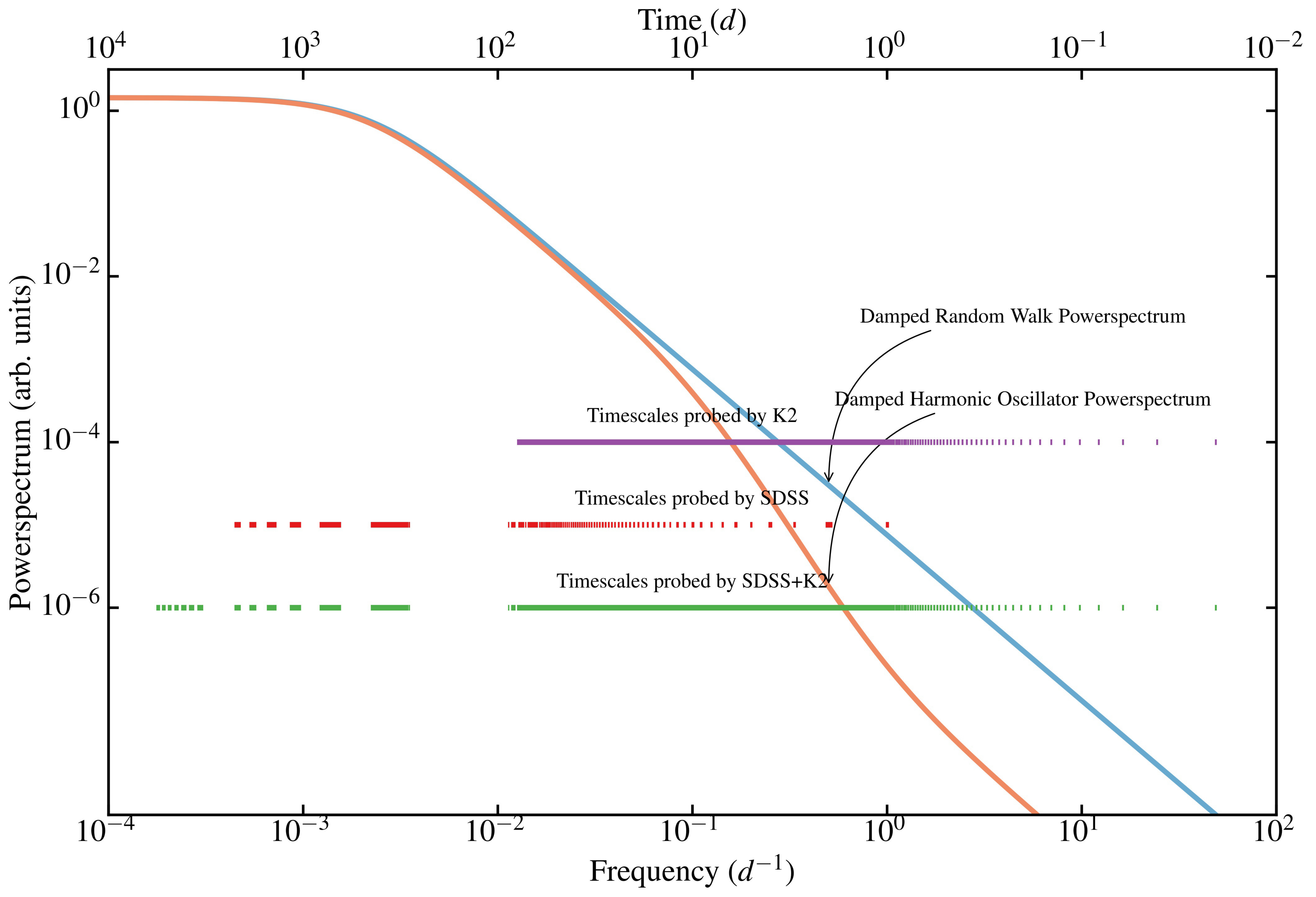}
    \caption{Why do AGN poorly-sampled AGN light curves look like a damped random walk (DRW)? We plot two PSD models - the DRW of \citet{Kelly09} in cyan and a damped harmonic oscillator (DHO) in orange. The DRW is a C-ARMA($1$,$0$) model i.e  the simplest possible C-ARMA model. The DHO is C-ARMA($2$,$q$) model, i.e., it is a $2^{\text{nd}}$-order C-ARMA model. The vertical dashes are located at the time-intervals probed by NASA's \textit{K2} mission (purple), the SDSS survey (red), and an idealized combination of the two (green). On timescales longer than $\sim 10$~d, both PSDs are identical, explaining why the simpler DRW model is popular for light curves that do not probe short timescales. The availability of better data in the future will necessitate more sophisticated C-ARMA models.}
    \label{fig:PowerOfSDSSK2}
\end{figure*}

We shall discuss the C-ARMA($2$,$1$) process in detail because it is relatively simple to work with and interpret. Second order differential equations also have strong physical motivations; $2^{\mathrm{nd}}$-order differential equations such as the wave equation govern many physical phenomena.

The process of linearising the original non-linear differential equation of equation~\eqref{eq:FluxDE} preserves the order of the original non-linear differential equation, i.e. no changes occur in the order of the highest derivative of the flux that appears in that equation. This means that if the original equation governing the flux emitted by the accretion disk is of $p^{\mathrm{th}}$-order, the corresponding linearised C-ARMA process is also of $p^{\mathrm{th}}$-order. 

The general form of the C-ARMA($2$,$1$) process is
\begin{equation}\label{eq:CARMA21}
\mathrm{d}^{2}\chi + \alpha_{1} \mathrm{d}\chi + \alpha_{2} \chi = \beta_{0}(\mathrm{d}W) + \beta_{1} \mathrm{d}(\mathrm{d}W),
\end{equation}
where $[\alpha_{1}] = T^{-1}$, $[\alpha_{2}] = T^{-2}$, $[\beta_{0}] = [\chi]~T^{-3/2}$, and $[\beta_{1}] = [\chi]~T^{-1/2}$. The LHS determines how individual perturbations evolve over time. The RHS determines the shape of the driving process PSD. If $\beta_{1} = 0$, the driving noise process PSD is flat over all frequencies, i.e., it is a white noise process. If $\beta \neq 0$, the disturbance process will have different amounts of power at different frequencies, i.e., it is a `colored' noise process. Equation~\eqref{eq:CARMA21} is of the same mathematical form as a \textit{Damped Harmonic Oscillator} (DHO) that is driven/perturbed by a colored noise process. Light curves that obey the C-ARMA($2$,$1$) process equation have perturbations from the mean flux level that individually behave like damped harmonic oscillators.
Given the identification of the  C-ARMA($2$,$1$) process with a DHO, $\alpha_{2} = \omega^{2}$ gives the angular frequency of the HO, and $\alpha_{1} = 2\omega\zeta_{\mathrm{dHO}}$ gives the damping ratio $\zeta_{\mathrm{dHO}}$ \citep[see chapter~7 for notation]{PanditWu}. The time-period of the harmonic oscillator is $T_{\mathrm{HO}} = 2 \upi/\omega$. The roots of the LHS of equation~\eqref{eq:CARMA21} are given by
\begin{equation}\label{eq:CARMA21Roots}
\rho_{1},\rho_{2} = -\omega\zeta_{\mathrm{dHO}} \pm \omega \sqrt{\zeta_{\mathrm{dHO}}^{2} - 1} = -\frac{\alpha_{1}}{2} \pm \sqrt{\frac{\alpha_{1}^{2}}{4} - \alpha_{2}}.
\end{equation}
The behavior of the flux perturbations is non-physical in this context if $\zeta_{\mathrm{dHO}} = 0$, i.e., the flux perturbations are undamped and never die out. For $0 < \zeta_{\mathrm{dHO}} < 1$, the flux perturbations are under-damped and will oscillate in value around the steady-state flux value with angular frequency $\omega' = \omega\sqrt{1 - \zeta_{\mathrm{dHO}}^{2}}$ but gradually decrease in amplitude. If $\zeta_{\mathrm{dHO}} = 1$, the perturbations are critically damped and will rapidly die down to the steady-state flux. If $\zeta_{\mathrm{dHO}} > 1$, flux perturbations are over damped and exponentially decay to the steady-state flux. The rate of decay decreases with $\zeta_{\mathrm{dHO}}$; over-damped flux perturbations with larger $\zeta$ take longer to decay, increasing the correlation timescale of the light curve. From equation~\eqref{eq:CARMAPSD}, the PSD of the C-ARMA($2$,$1$) process is given by
\begin{equation}\label{eq:CARMA21PSD}
S_{\chi\chi}(\nu) = \frac{1}{2\upi}\frac{\beta_{0}^{2} + 4\upi^{2}\beta_{1}^{2}\nu^{2}}{16\upi^{4}\nu^{4} + 4\upi^{2}(\alpha_{1}^{2}-2\alpha_{2})\nu^{2} + \alpha_{2}^{2}},
\end{equation}
which has frequency-dependent behavior.

Figure~\ref{fig:PowerOfSDSSK2} shows the dependence of a DHO PSD on frequency. At very high- and low-frequencies, the DHO PSD looks like the damped random walk PSD of \citet{Kelly09} with $S_{\chi\chi} \sim 1/\nu^{2}$. There is a mid-range of frequencies where the $\nu^{4}$ term in the denominator dominates. At the lowest frequencies, the PSD is dominated by the frequency independent terms and asymptotically approachs $\beta_{2}^{2}/2\upi \alpha_{2}^{2}$. This frequency-dependent PSD behavior may explain why ground-based observations of AGN light curves suggest that AGN variability is well-modelled by a DRW whereas \textit{Kepler} observations of AGN suggest that the PSD is steeper than a DRW: the ground based observations are not sampling at high enough frequency to be able to see the $\nu^{4}$ dominated regime.

Individual accretion disk perturbations are smeared by processes such as turbulent dissipation, rotation and thermal dissipation. The nature of these processes can be inferred via the Green's function approach as discussed in section~\ref{sec:Meaning}. The Green's function of the DHO is given by
\begin{equation}\label{eq:DHOGFunc}
G(t) = \frac{\mathrm{e}^{\rho_{1}t} - \mathrm{e}^{\rho_{2}t}}{\rho_{1} - \rho_{2}},
\end{equation}
where $\rho_{1}$ and $\rho_{2}$ are the roots of the autoregressive characteristic polynomial of equation~\eqref{eq:CARMA21Roots}. If $\zeta_{\mathrm{dHO}} > 1$, i.e., flux perturbations are constrained to be damped out by the dissipative processes in the accretion disk, then the roots are purely real-valued. A perturbation to the accretion disk flux in the form of an impulse at $t = t_{0}$ will result in the flux increasing from $t_{0}$ to $t_{max}$ given by
\begin{equation}\label{eq:maxT}
t_{\mathrm{max}} = \frac{1}{\rho_{2} - \rho_{1}} \ln \left| \frac{\rho_{1}}{\rho_{2}} \right|.
\end{equation}
After $t_{\mathrm{max}}$, the effect of the original perturbation decays to the steady-state flux level. Although the decay is not a pure exponential, it is useful to characterize the decay rate by an e-folding time-scale $t_{\mathrm{e-fold}}$. It may be computed by solving
\begin{equation}\label{eq:TEFold}
G(t_{\mathrm{e-fold}}) = \frac{G(t_{\mathrm{max}})}{\mathrm{e}},
\end{equation}
numerically for $t_{\mathrm{e-fold}}$.

From the PSD of the C-ARMA($2$,$1$) process in equation~\eqref{eq:CARMA21PSD}, we see that on very long time-scales, both the numerator and the denominator should approach a constant value. The PSD begins to flatten on time-scale $t_{\mathrm{flat}}$ where
\begin{equation}\label{eq:TFLAT}
t_{\mathrm{flat}} \sim \max \left( \frac{2\upi \beta_{1}}{\beta_{0}},\frac{2\upi}{\sqrt{\alpha_{2}}},\frac{2\upi \sqrt{\alpha_{1}^{2}-2\alpha_{2}}}{\alpha_{2}} \right).
\end{equation}
On the other hand, on very short time-scales, the $\nu^{4}$ term in the denominator dominates. The PSD $\sim \nu^{-4}$ on time-scale $t_{\mathrm{steep}}$ given by
\begin{equation}\label{eq:STEEP}
t_{\mathrm{steep}} \sim \min \left( \frac{2\upi}{\sqrt{\alpha_{1}^{2} - 2\alpha_{2}}}, \frac{2\upi}{\sqrt{\alpha_{2}}}  \right).
\end{equation}

We see that the C-ARMA($2$,$1$) model contains several time-scales of interest: the peak of the Green's function($t_{\mathrm{max}}$), the e-folding time-scale of the Green's function ($t_{\mathrm{e-fold}}$), the time-scale on which the PSD of the driving flux disturbances turns over ($t_{\mathrm{turn}}$), the time-scale on which the PSD begins to flatten ($t_{\mathrm{flat}}$), and finally the time-scale on which the PSD $\sim \nu^{4}$ ($t_{\mathrm{steep}}$). All of these time-scales should be detectable in the PSD of the light curve in the form of various bends and turnovers. Similar time-scales can easily be derived for higher-order C-ARMA models with the number of significant time-scales increasing with the model order, i.e., more complex C-ARMA models will have more `features' in the PSD indicative of still more complex processes at work in the accretion disk.

In the case of the C-ARMA($2$,$1$) process, the driving noise is best interpreted as the `acceleration' of flux perturbations away from the steady-state flux level. The PSD of the driving noise process is given by
\begin{equation}\label{eq:CARMA21distPSD}
S_{uu}(\nu) = \frac{1}{2 \upi}(\beta_{0}^{2} + 4 \upi^{2} \beta_{1}^{2} \nu^{2}),
\end{equation}
and can be described (at higher frequencies) as a violet- or purple-noise spectrum. At low frequencies, the $\beta_{0}^{2}$ term in the RHS dominates: the log-PSD slope is $0$ and the PSD is similar to that of white-noise. The transition from the $\beta_{0}^{2}$-dominated to the $4 \upi^{2} \beta_{1}^{2} \nu^{2}$-dominated behaviour occurs at
\begin{equation}\label{eq:distPSDTurnover}
t_{\mathrm{turn}} = \frac{2 \upi \beta_{1}}{\beta_{0}}.
\end{equation}
Thermal motion within a fluid results in the formation of sound-waves in the fluid with characteristic PSD $\sim \nu^{{2}}$ \citep{Mellen52} suggesting that the driving noise PSD of CARMA($2$,$1$) models can be ascribed to the thermal noise of the accretion disk material. Alternatively, it may be caused by the presence of eddies in the turbulent flow of the accretion disk \citep{EddySim15}.

In the next section, we discuss how to infer the values of the coefficients in the C-ARMA process describing accretion disk fluctuations in an AGN from observations of the light curve of the object.

\section[\href{https://github.com/AstroVPK/kali}{\textsc{k\={a}l\={i}}}]{\href{https://github.com/AstroVPK/kali}{\textsc{k\={a}l\={i}}}: Software for the C-ARMA Analysis of a Light Curve}\label{sec:Steps}

We have implemented a fully parallelized and vectorized software package, \href{https://github.com/AstroVPK/kali}{\textsc{k\={a}l\={i}}} (named jointly after the Hindu goddess of time, change, and power and also as an acronym for KArma LIbrary), to analyze stochastic light curves using the C-ARMA process presented in equation~\eqref{eq:CARMAIntro}. \href{https://github.com/AstroVPK/kali}{\textsc{k\={a}l\={i}}} is implemented in the \textsc{c++} programming language with \textsc{Python} language bindings for ease of use. \href{https://github.com/AstroVPK/kali}{\textsc{k\={a}l\={i}}} may be obtained from \url{https://github.com/AstroVPK/libcarma} and is installable on Linux variants and Mac OSX using the install scripts provided. Appendix~\ref{sec:kali} describes the mathematics implemented in \href{https://github.com/AstroVPK/kali}{\textsc{k\={a}l\={i}}} that are used to perform the Bayesian Markov Chain Monte-Carlo (MCMC) inferencing of a stochastic light curve.

Before performing a C-ARMA analysis, it should first be determined if a C-ARMA process is an appropriate model for the light curve of the AGN. A C-ARMA process may be unsuitable if the light curve exhibits non-stationary behaviour, i.e., (1) the powerspectrum of the light curve changes over the course of the light curve at the $2$-$\sigma$ level as determined by estimating the powerspectrum of short segments of the light curve; or (2) the light curve has a marked linear trend as determined from a simple regression test. A C-ARMA analysis may also be unsuitable if the light curve exhibits sudden, large amplitude-short duration changes in flux i.e. `flares' (as in the case of the blazar of \citealp{Edelson13}). A C-ARMA process would have to generate a series of positive large amplitude variations followed by a similar series of negative variations to produce `flare'-like behavior which is mathematically highly unlikely to occur. C-ARMA processes are theoretically able to produce any value for the flux variation, including values that would result in \textit{negative} total flux. Therefore, C-ARMA processes are unsuitable models for AGN variability if the flux variations are very large compared to the flux as this may increases the likelihood of producing an unphysical negative total flux to an unacceptable level. A C-ARMA model is suitable if the sample autocorrelation and partial autocorrelation functions of the light curve exponentially decay to below the $2$-$\sigma$ significance level for a pure white noise process \citep{BrockwellDavisITSF}.

Once it is decided that a C-ARMA process may be an appropriate choice of model for the light curve, we must estimate the model-order of the C-ARMA process. Some insight may be available from the autocorrelation and partial autocorrelation functions (see \citealp{BrockwellDavisITSF} for details). We automate the process by sampling parameters from a range of models with increasing numbers of C-ARMA parameters and using the \textit{Deviance Information Criteria} (DIC) of equation~\eqref{eq:DIC} to select the best model. For a given model order $p$ and $q$, we use MCMC to sample the space of model parameters. Before beginning the Kalman recursions of appendix~\ref{sec:Kalman}, we construct a mask matrix to keep track of missing observations and subtract the mean of the light curve from every observation to make the light curve a zero mean process. For a given set of model parameters ($\alpha_{k}$ and $\beta_{k}$), we check for the validity of the parameter set, i.e., are real parts of the roots of the autoregressive and moving average polynomials less than zero? If they are not, we reject the model outright by setting the prior likelihood to $0$. If the model parameters are permissible, we set the prior likelihood to $1$ and compute the transition matrix $\mathbfss{F}$ using equation~\eqref{eq:TransMat} and the disturbance variance-covariance matrix $\mathbfss{Q}$ using equation~\eqref{eq:DistMat}. We then use equations~\eqref{eq:x0} and~\eqref{eq:P0} to initialize the state $\widehat{\mathbfit{x}}^{+}_{0}$ and determine the initial state uncertainty $\mathbfss{P}^{+}_{0}$. To compute the likelihood of the light curve, we iterate through the observations using the Kalman filtering equations~\eqref{eq:APrioriState} through~\eqref{eq:APosterioriUncertainty} to compute the innovation and the uncertainty of the innovation for every observation. The computed innovations and uncertainties are used to calculate the likelihood using equation~\eqref{eq:DataGivenModel} which is then equal to the model likelihood of equation~\eqref{eq:ModelGivenData} given our choice of the prior. The MCMC algorithm can then sample the parameter space and obtain draws of the C-ARMA model parameters from the posterior probability distribution of the parameters. Using the DIC with the full set of draws (Bayesian approach), or the \textit{corrected Akaike Information Criteria} (AICc) with the parameter set that has the highest likelihood (frequentist approach), we can pick the best fitting values of $p$ and $q$ simultaneously with the parameter estimation process. Once we have selected the model order, we can use the inferred parameter values to compute estimates of (1) the Green's function; (2) the PSD of the driving impulses; and (3) time-scales such as $t_\mathrm{max}$.

In the next section, we present a study of the accretion physics of the Seyfert 1 AGN Zw 229-15 using data from the \Kepler mission.

\section[C-ARMA Analysis of Zw 229-15]{Probing Accretion Physics with the \Kepler Seyfert 1 Zw 229-15: C-ARMA Analysis of the Light Curve}\label{sec:Zw229-15}

We apply the C-ARMA inferencing techniques described in sections~\ref{sec:Fitting} through~\ref{sec:ModelSelection} to the re-processed \Kepler light curve of the Seyfert 1 AGN \href{http://ned.ipac.caltech.edu/cgi-bin/objsearch?objname=Zw+229-15&extend=no&hconst=73&omegam=0.27&omegav=0.73&corr_z=1&out_csys=Equatorial&out_equinox=J2000.0&obj_sort=RA+or+Longitude&of=pre_text&zv_breaker=30000.0&list_limit=5&img_stamp=YES}{Zw 229-15} (\Kepler catalog name KIC 006932990) located at $z = 0.028$. Details of the re-processing applied to the light curve can be found in \citep{CariniWilliamsAAS} and in \citet{Kasliwal15b}.

\subsection[C-ARMA Model for Zw 229-15]{Determination of a C-ARMA Model for the Light Curve of Zw 229-15}\label{sec:Zw229-15CARMA}

\begin{figure*}
    \includegraphics[width=\textwidth]{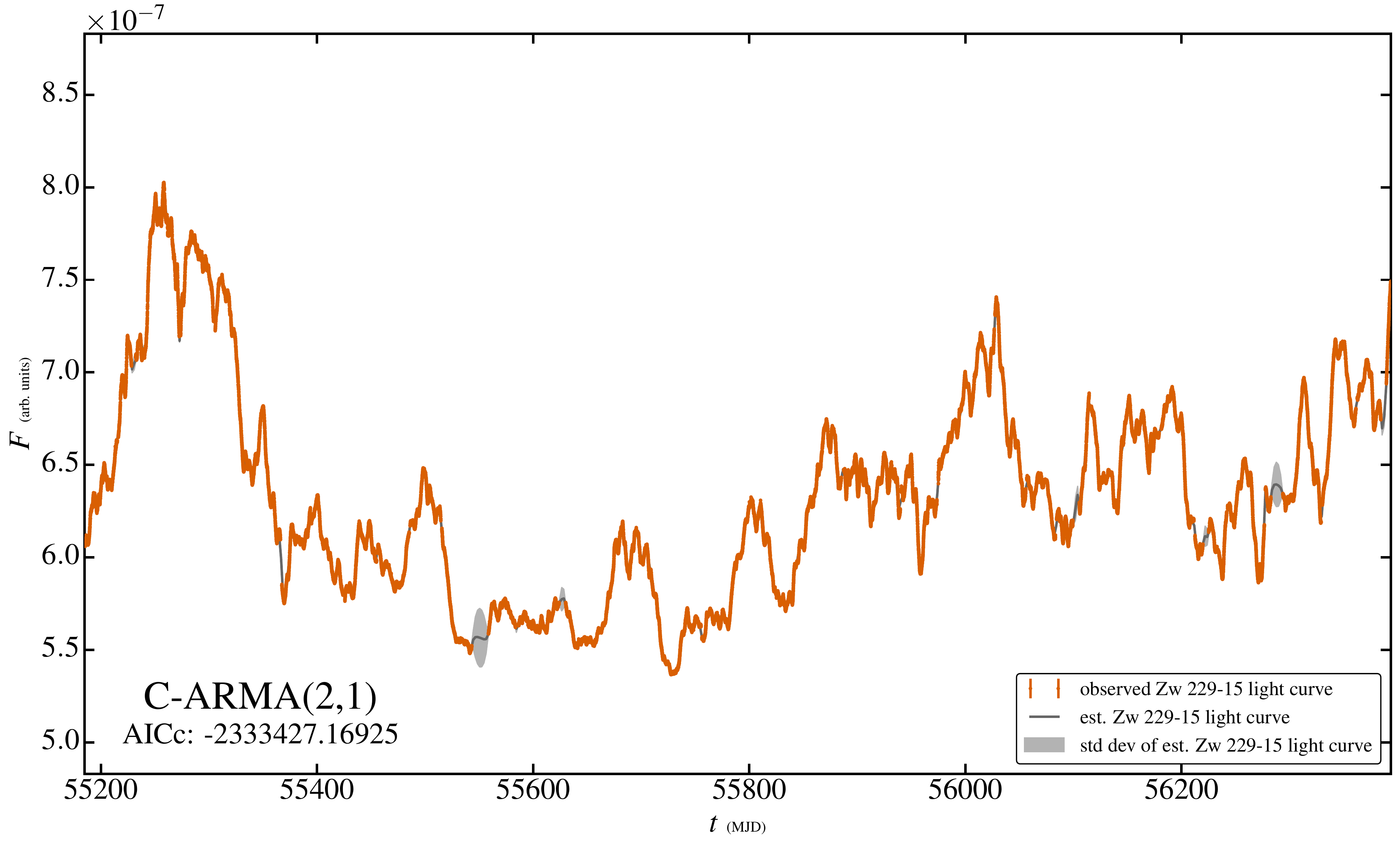}
    \caption{Light curve of the Seyfert 1 AGN Zw 229-15. This light curve is best fit by the C-ARMA($2$,$1$) process discussed in section~\ref{sec:CARMA21}.}
    \label{fig:Zw229-15_LC}
\end{figure*}

We use our package, \href{https://github.com/AstroVPK/kali}{\textsc{k\={a}l\={i}}} to fit the observed light curve of Zw 229-15 to C-ARMA($p$,$q$) processes with $1 \leq p \leq 10$ and $0 \leq q < p$. In addition to using our package \href{https://github.com/AstroVPK/kali}{\textsc{k\={a}l\={i}}}, we performed the same analysis using the \textsc{c++} and \textsc{Python} library \textsc{carma\_pack} of \citet{Kelly14}. While both codes produced similar numerical results, \href{https://github.com/AstroVPK/kali}{\textsc{k\={a}l\={i}}} runs about $50 \times$ faster due to our extensive usage of the high performance Intel Math Kernel Library (MKL). Prior to the analysis, the light curve of Zw 229-15 was corrected to the rest-frame of the galaxy at $z = 0.0275$ by scaling $\delta t$ by $1/(1+z)$. No binning of the light curve was performed. Instead of adopting the \Kepler error estimates, \textsc{carma\_pack} assumes that the variance of the observation noise may be inaccurate and sets $\sigma^{2}_{N,\mathrm{true}} = M_{\mathrm{Err}} \sigma^{2}_{N,\mathrm{stated}}$ for each observation. The multiplicative scaling factor $M_{\mathrm{Err}}$ is treated as a fit parameter, i.e., all purely white noise in the data is assumed to originate in process of observation rather than be intrinsic to the signal. We find that the median value of the multiplicative scaling factor is $M_{\mathrm{Err}} = 1.30$. The observation noise levels inferred from the pixel noise statistics of the \Kepler CCDs appear to underestimate the true observation noise level in the signal by $\sim 14$ percent at the $95$-percent confidence level.

\begin{figure*}
    \includegraphics[width=\textwidth]{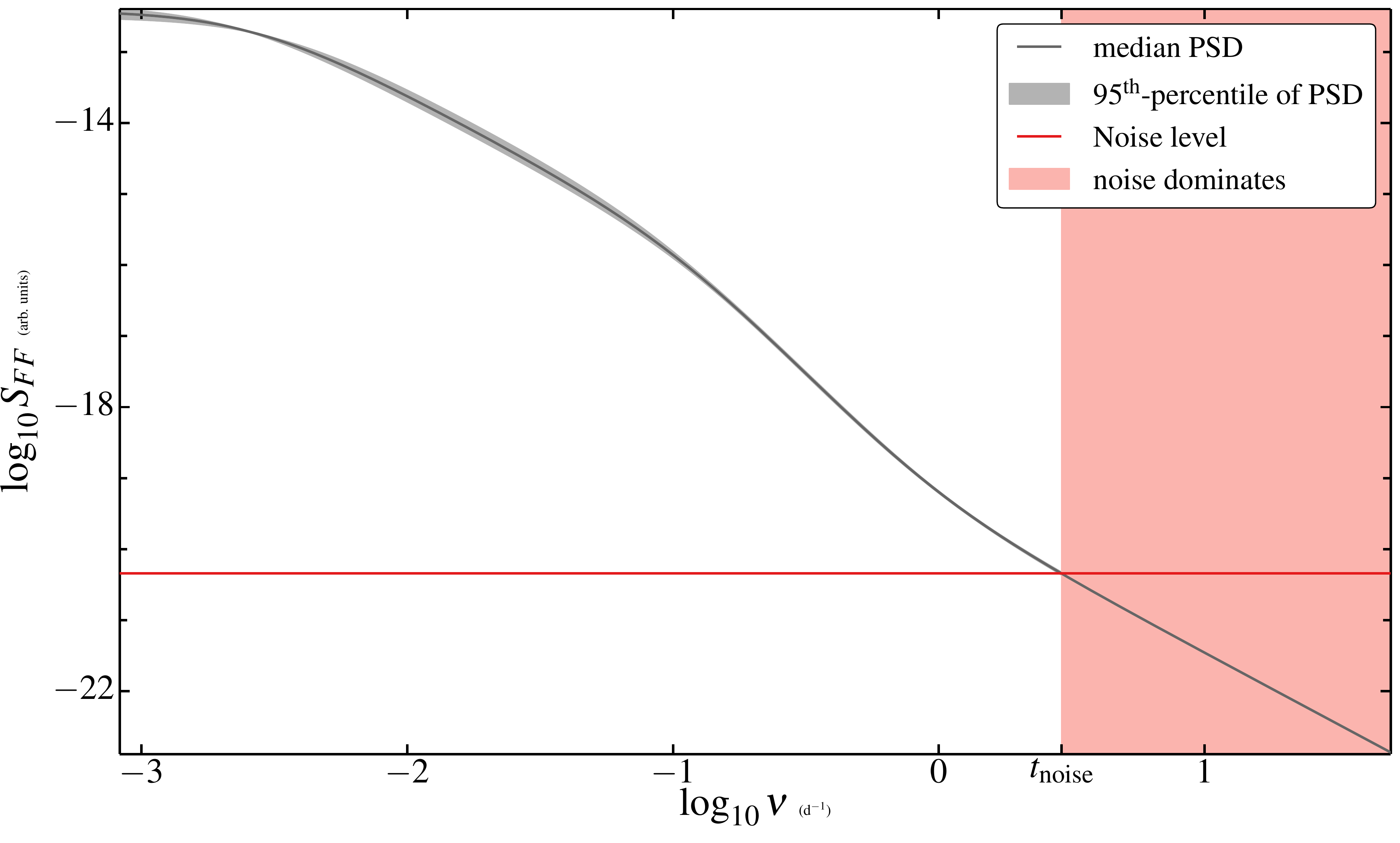}
    \caption{Power spectral density of the light curve of Zw 229-15. Using the \Kepler noise properties $\sigma^{2}_{N,\mathrm{stated}}$ and $M_{\mathrm{Err}}$, we compute the observation noise level (red line). The inferred PSD crosses the noise level at $t_{\mathrm{noise}} \sim 8.3$~h. On time-scales comparable to $t_{\mathrm{noise}}$, i.e., $< 1$~d, the noise level does not permit us to infer anything about the behaviour of this light curve. Compare this PSD to the DHO PSD in figure~\ref{fig:PowerOfSDSSK2}.}
    \label{fig:Zw229-15_PSD}
\end{figure*}

Although we searched an extensive space of C-ARMA models ($1 \leq p \leq 10$ and $0 \leq q < p$), the relatively simple C-ARMA($2$,$1$) model achieved the lowest AICc value, i.e., the best fit. The AICc value of $-2333427.16925$ for the C-ARMA($2$,$1$) model v/s the value $-2333328.70173$ for the C-ARMA($3$,$2$) model (the next best model) suggests that the C-ARMA($3$,$2$) is very unlikely ($\sim 10^{-22}$ times as likely) to explain the observed data \citep{ModelSelection}. Figure~\ref{fig:Zw229-15_LC} shows the full light curve of Zw 229-15 (orange) along with the RTS smoothed realization (grey) of section~\ref{sec:Smoothing}. In addition to the expectation value of the smoothed realization, we show (light grey) the $1\sigma$ limits on the smoothed light curve. Notice how this $1\sigma$ limit is usually too small to be visible but increases dramatically when the observed light curve has missing values. Figure~\ref{fig:Zw229-15_PSD} shows the median estimated PSD of the light curve of Zw 229-15 (grey) along with the $95^{\mathrm{th}}$-percentile confidence region (light grey). From the reported \Kepler noise estimates and the estimates of $M_{\mathrm{Err}}$, we compute the noise level in the PSD using $S_{NN} = 2\delta t M_{\mathrm{Err}} \sigma^{2}_{N}$ where we use median values for $M_{\mathrm{Err}}$ and $\sigma^{2}_{N}$. We estimate the frequency $\nu_{\mathrm{noise}} \sim 2.9$~d$^{-1}$ at which the PSD of the light curve drops below the noise level by linearly interpolating the PSD between points. The corresponding time-scale is $t_{\mathrm{noise}} \sim 8.3$~h. Since the PSD of the underlying flux variations drops below the PSD of the \Kepler noise level at $t_{\mathrm{noise}}$, we have no information about the behaviour of the light curve on time-scales shorter than $\sim 8$~h and any features that appear on time-scales of under $\sim 16$~h should be regarded as dubious given the noise properties of \Kepler. We mark the noise-dominated region of the PSD plot in the inset in red.

\subsection[dHO Behavior of Zw 229-15]{The Damped Harmonic Oscillator-like Behavior of the Light Curve of Zw 229-15}\label{sec:DHO}

\begin{figure}
    \includegraphics[width=\columnwidth]{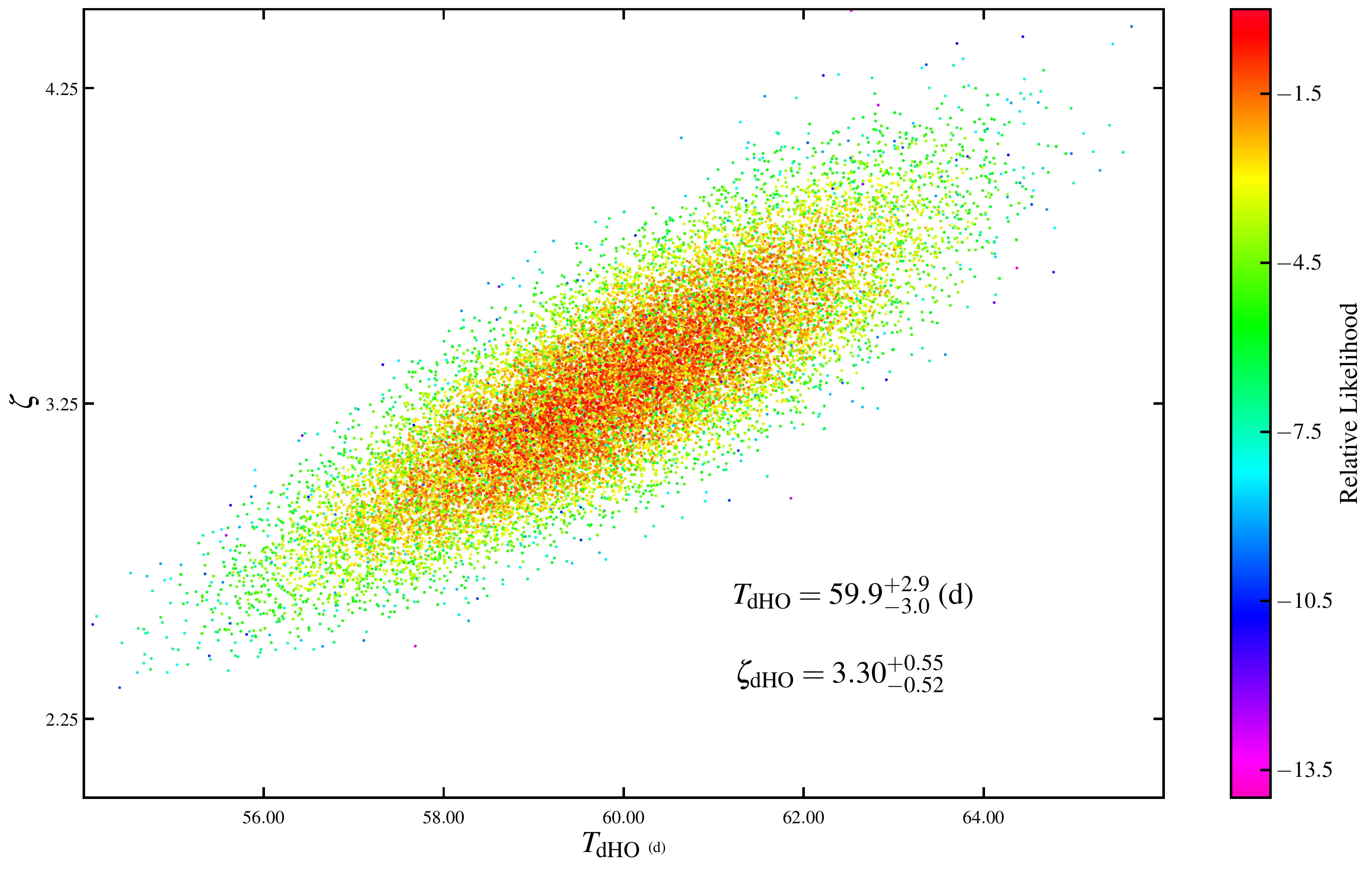}
    \caption{Damped Harmonic oscillator parametrization of the light curve of Zw 229-15. The time period of the oscillator is $T$ and the damping ratio is $\zeta$. The perturbations in the accretion disk of Zw 229-15 are over-damped as indicated by the mean value of $\zeta > 1$. This indicates that flux perturbations steadily decay to the mean flux level without oscillating.}
    \label{fig:Zw229-15_HO}
\end{figure}

Since the Zw 229-15 light curve is well-described by a C-ARMA($2$,$1$) process, the differential equation governing the behavior of flux perturbations may be interpreted as the damped harmonic oscillator driven by a colored stochastic process as described in section~\ref{sec:CARMA21}. Figure~\ref{fig:Zw229-15_HO} shows the distribution of the un-damped-oscillator time period $T_{\mathrm{dHO}}$ against the damping ratio $\zeta_{\mathrm{dHO}}$. The damping ratio is very large with median value $3.30$ with $95$-percent CI [$2.78$,$3.85$] while the time period of the harmonic oscillator in the absence of damping is $59.9$~d with $95$-percent CI [$56.9$,$62.8$]~d. Physically this implies that flux perturbations slowly return to the steady-state flux level after reaching peak intensity rather than oscillating about the mean flux level. This observation supports the idea that flux perturbations may occur due to local Magneto Rotational Instability-generated hot- and cold-spots in the accretion disk that gradually disperse over time.

\begin{figure*}
    \includegraphics[width=\textwidth]{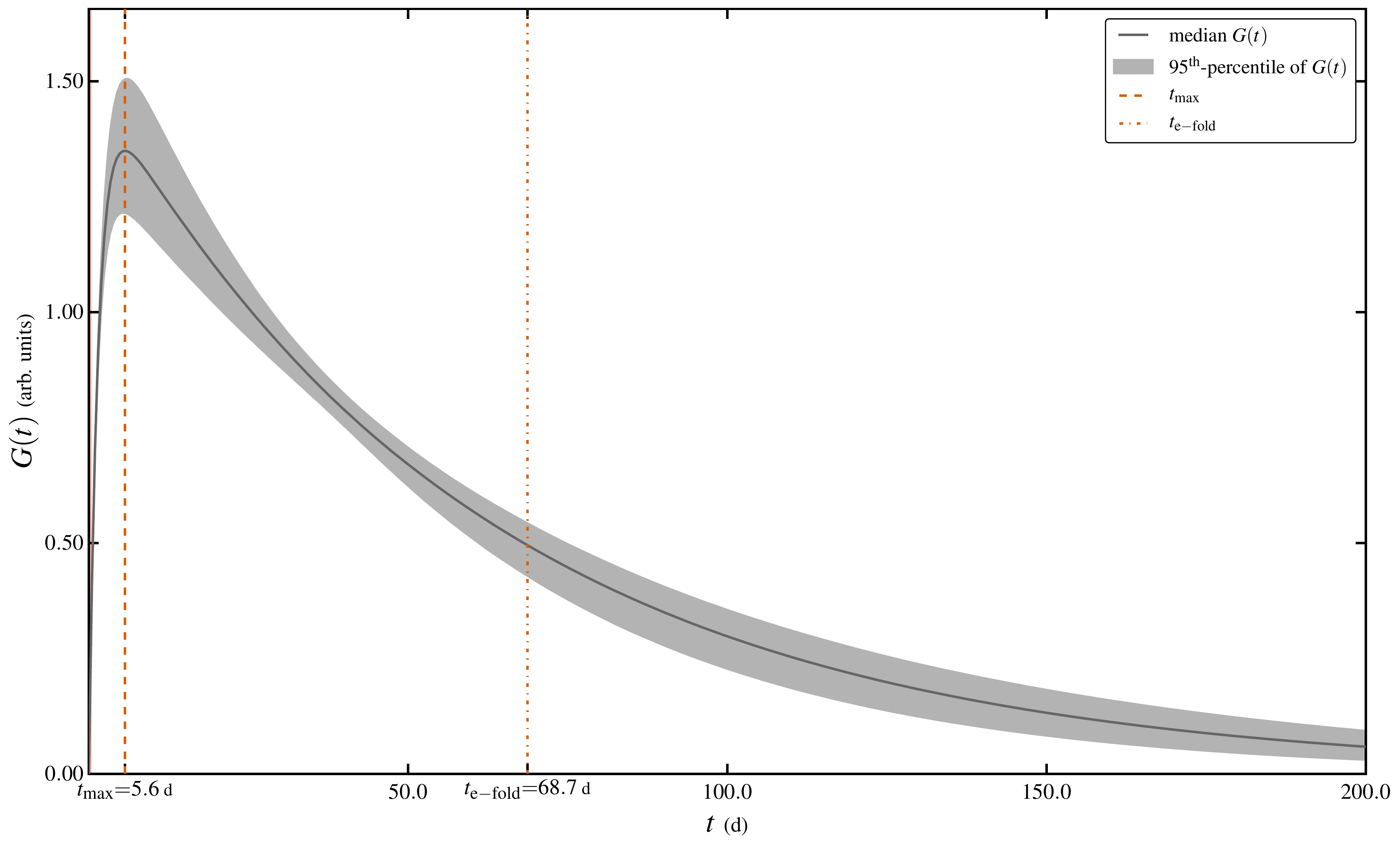}
    \caption{The Green's function for the light curve of Zw 229-15 quantifies how a unit impulse flux perturbation evolves as a function of time. A tiny unit impulse perturbation at $t = 0$ causes the flux from the perturbation to peak at $t = t_{\mathrm{max}}$ ($\sim 5.65$~d). The flux perturbation then decays exponentially with $t_{\mathrm{e-fold}} \sim 68.7$~d, dropping to under $\sim 1$ percent the peak intensity $\sim 320$~d later.}
    \label{fig:Zw229-15_GFunc}
\end{figure*}

The corresponding Green's function is shown in figure~\ref{fig:Zw229-15_GFunc} along with the $95$ percent CI. Recall that the Green's function quantifies the evolution of a `unit-impulse'. In this case, the impulses consist of perturbations of the flux away from the steady-state level. The effect of an impulse is to drive the flux away from the mean-level until the rate at which the flux is changing drops to zero. This occurs $t_{\mathrm{max}} = 5.65$~d with $95$-percent CI [$5.50$,$5.80$]~d after the original impulse (orange dashed line). On the left hand side of the figure, we see that on short time-scales (red-shaded region) the observation noise makes it impossible to observe structure in the Green's function. However, the $t_{\mathrm{max}} \sim 5.6$~d time-scale that we detect here is safely above $t_{\mathrm{noise}} \sim 8.3$~h and is therefore unlikely to be an artifact of the instrumentation-noise. Due to the very large damping ratio, flux perturbations do not dissipate rapidly---perturbations drop in intensity from the peak by a factor of $\mathrm{e}$ by $t_{\mathrm{e-fold}} = 68.7$~d with $95$-percent CI [$62.38$,$75.08$]~d. On longer time-scales, the correlation between the total integrated flux from the accretion disk and the original perturbation will decrease as the flux becomes dominated by the increasingly larger numbers of intervening perturbations. It is plausible that conventional approaches such as PSD-fitting and structure function analysis detect the $t_{\mathrm{e-fold}}$ time-scale as the de-correlation time-scale.

\begin{figure}
    \includegraphics[width=\columnwidth]{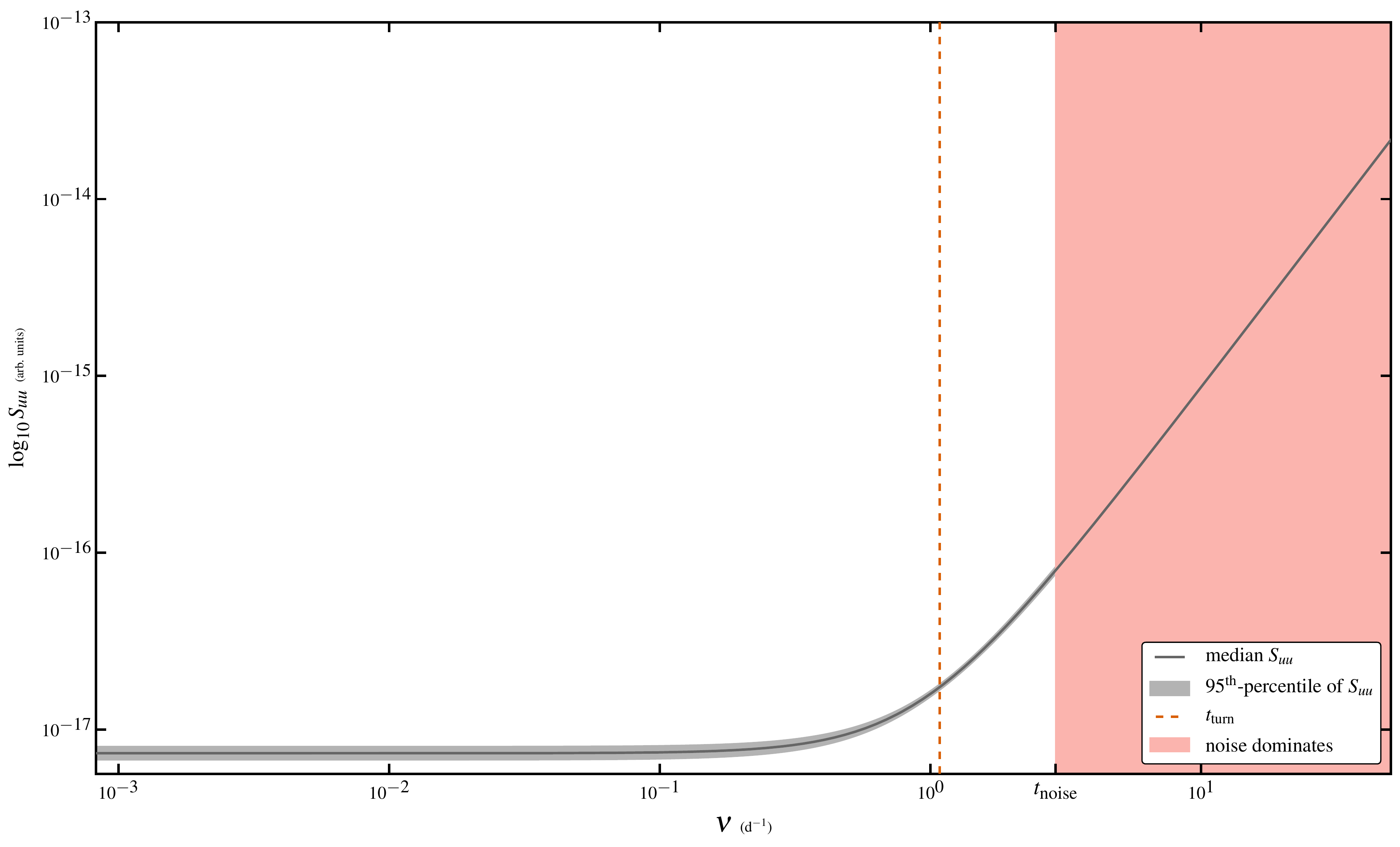}
    \caption{PSD of the flux perturbation impulses in the accretion disk of Zw 229-15. The PSD begins to steepen at $t = t_{\mathrm{turn}} (\sim 1)$~d i.e on time-scales longer than $\sim 1$~d, the log PSD has no slope implying that the flux perturbation impulses behave like white noise. On shorter time-scales, the log PSD increases linearly with slope $2$.}
    \label{fig:Zw229-15_distPSD}
\end{figure}

Figure~\ref{fig:Zw229-15_distPSD} shows the PSD of the disturbance properties inferred from the estimates of $\beta_{1}$ and $\beta_{2}$, i.e., the powerspectrum of the impulses that drive the flux perturbations. On very long time-scales, the disturbance have equal power over $\sim 3.5$ decades of frequency. On time-scales of $\sim 1$~d, the disturbance PSD begins to rise with log-slope $2$. Impulses with such PSD resemble `violet noise'. Such PSD are produced by the thermal noise of the medium \citep{Mellen52} suggesting that on these very short time-scales, we may be probing the accretion disk matter directly. It may also be the case that we are detecting very low wavenumber, large lengthscale eddies in the turbulent flow \citep{EddySim15}. Unfortunately, the proximity of the inferred turnover time-scale $t_{\mathrm{turn}} = 1.081$~d with $95$-percent CI [$1.05$,$1.11$]~d to the noise-dominated time-scale $t_{\mathrm{noise}} \sim 8.3$~h makes it difficult to draw conclusions about the disturbance PSD behaviour on these time-scales.

\begin{figure*}
    \includegraphics[width=\textwidth]{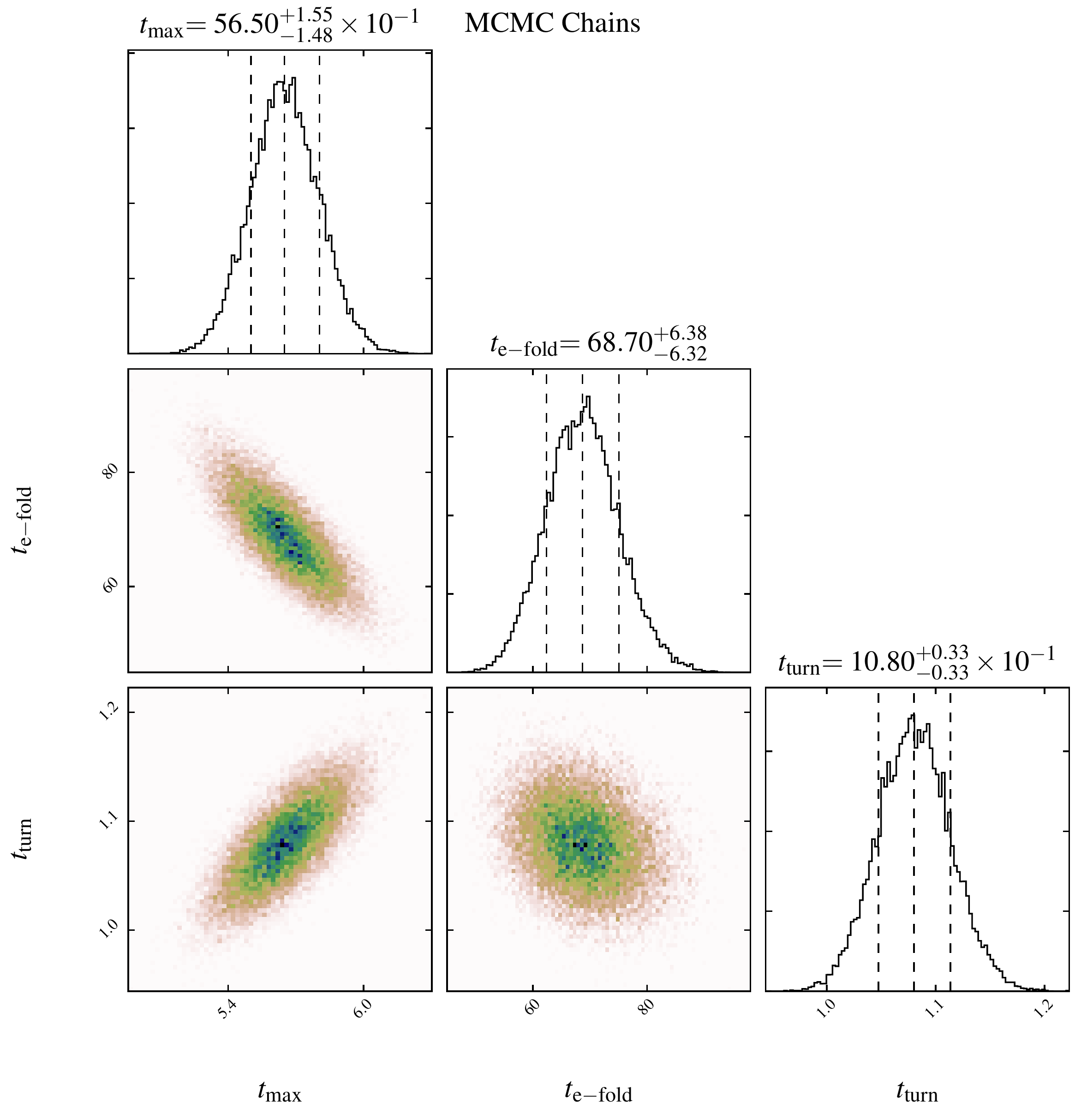}
    \caption{Scatter plots of $t_{\mathrm{max}}$ v/s $t_{\mathrm{e-fold}}$ v/s $t_{\mathrm{turn}}$ in units of d. The diagonal shows histograms of the estimates of these time-scales. The time taken to achieve peak output flux, $t_{\mathrm{max}}$, is anti-correlated with the e-folding time of the flux perturbation $t_{\mathrm{e-fold}}$ but correlated with $t_{\mathrm{turn}}$.}
    \label{fig:Zw229-15_times}
\end{figure*}

Using equation~\eqref{eq:TFLAT}, the PSD of the Zw 229-15 light curve flattens on time-scales greater than $t_{\mathrm{flat}} = 386.78$~d with $95$-percent CI [$308.36$,$472.36$]~d. Evidence of this flattening is barely visible at the lowest frequencies in figure~\ref{fig:Zw229-15_PSD}. A visual indication of the flattening of the PSD would require that the sampled light curve be more than $\sim 10 \times$ the detected flattening time-scale. Is it possible to determine the optimal duration to sample the light curve in real-time, i.e., in the middle of obtaining the observations? During the course of the observations, we can use the Kalman filter to analyse the existing data and update the existing best estimate of  $t_{\mathrm{flat}}$ as each new observation is made. Our prediction is that if the light curve is stationary, the estimates of $t_{\mathrm{flat}}$ will eventually converge to the true value. At this point we would have an estimate of the duration that we should sample the light curve if we desire to see the PSD flatten from the observations themselves. We also estimate $t_{\mathrm{steep}} = 9.29$~d with $95$-percent CI [$8.20$,$10.66$]~d. This is \textit{longer} than $t_{\mathrm{turn}}$---we see that the effect of the moving average parameter $\beta_{1}$ is to introduce short time-scale power into the light curve on time-scales under $\sim 1$~d. Figure~\ref{fig:Zw229-15_times} shows cross-correlation plots of the various time-scales. Note that the estimates of $t_\mathrm{e-fold}$ and $t_{\mathrm{flat}}$ are poorly constrained. Future time-domain surveys such as LSST will probe a much longer temporal baseline. This should help constrain both $t_{\mathrm{e-fold}}$ and $t_{\mathrm{flat}}$ more tightly.



The light curve of Zw 229-15 observed by \Kepler is well described by a C-ARMA($2$,$1$) process suggesting that a $2^{\mathrm{nd}}$-order differential equation may be responsible for smoothing out flux perturbations over time. The \Kepler data suggest that the impulses to the flux that drive the perturbations have equal power on time-scales ranging from days to years. Perturbations to the flux behave like over-damped harmonic oscillators---the perturbation peaks in intensity at $\sim 5.6$~d after which it gradually decays in intensity with an $\mathrm{e}$-folding time-scale of $\sim 69$~d. On very short time-scales ($\sim 1$~d) there are indications that the impulse powerspectrum begins to rise with log-PSD slope $2$ which may be due to the thermal motion of the accretion disk plasma. However, the observation noise level makes it difficult to probe accretion physics on these time-scales. We discuss the effect of the \Kepler noise characteristics on our results in the next section.

\subsection[Moire Pattern Noise]{Moire Pattern Noise in the Light Curve of Zw 299-15}\label{sec:MPD}

Instrument-induced artifacts were expected in the \Kepler data prior to launch based on extensive ground-testing \citep{Kol10}. Further testing was performed during the commissioning of the telescope. Based on the testing, it was determined that several artifacts did not require mitigation \textit{for the primary science mission of} \Kepler: exo-planet detection. \citet{Kol10} recommended strategies for correcting or flagging several of the artifacts generated by the \Kepler detector electronics. Some of these recommendations have been implemented in the new \textit{Dynablack} module \citep{Clarke14} that has been added to the \Kepler SOC pipeline for Data Release 24 \citep{DR24}. The light curve that we analyze here is \textit{not} created using the per-pixel data provided by DR24--therefore we perform a test to quantify the effects of the various \Kepler detector-induced artifacts on the results presented in section~\ref{sec:DHO}.

The \Kepler focal plane science CCDs are read out over $84$ channels that are referred to using the module\#.output\# convention presented in \citet{KIH} and in \citet{Kasliwal15}. Zw 229-15 lands on the module\#.output\# combinations $14$.$48$, $8$.$24$, $12$.$40$, and $18$.$64$ as the \Kepler focal plane performs a $90$ degree rotation after every quarter, i.e., the first quarter of Zw 229-15 data, obtained during quarter $4$ (Q$4$) of the \Kepler mission were read out over $14$.$48$. Zw 229-15 landed on $14$.$48$ again during quarters Q$8$, Q$12$, and Q$16$. Of these module\#.output\# combinations, 14.48 exhibits moderate levels of rolling band and moire artifacts while 12.40 has an out of specification undershoot--see \citet{KIH} for a detailed explanation of the sources of these artifacts. We assess the impact of the rolling-band and moire artifacts by removing the affected quarters (Q$4$, Q$8$, Q$12$, and Q$16$) from the light curve and re-doing the C-ARMA model fit.

We find that the C-ARMA($2$,$1$) model still provides the best AICc value, i.e., the order of the fit was not sensitive to the moire and rolling band issues. The inferred excess instrumentation noise is $M_{\mathrm{Err}} \sim 1.24$ which implies that the measurement errors are $\sim 11.3$ percent higher than the quoted value. This suggests that there is a systematic underestimation of the \Kepler observation noise level that cannot be attributed to the abnormal channel. From the inferred observation noise level, we find that the noise confusion limit drops slightly to $t_{\mathrm{noise}} = 7.9$~h. The largest change is manifest in the value of the damping ratio of the damped harmonic oscillator $\zeta_{\mathrm{dHO}} = 2.97$ with $95$-percent CI [$2.40$,$3.57$] while the time period of the harmonic oscillator remains $T_{\mathrm{dHO}} = 59.0$~d with $95$-percent CI [$55.4$,$62.6$]~d. The Green's function peaks $t_{\mathrm{max}} = 5.89$~d after the initial perturbing impulse with $95$-percent CI [$5.54$,$6.26$]~d, after which it begins to slowly decay with an e-folding time of $t_{\mathrm{e-fold}} = 61.7$~d with $95$-percent CI [$48.8$,$76.0$]~d. The turnover time-scale on which the driving impulse PSD is observed to begin rising with log-PSD slope $2$ occurs at $t_{\mathrm{turn}} = 1.16$~d with $95$-percent CI [$1.08$,$1.24$]~d.

All of these estimates are consistent (to within the $95$-percent confidence intervals computed) with the corresponding values for the full light curve. This suggests that while the data obtained over module and output $14$.$48$ have higher levels of observation noise and spacecraft-induced systematics, the observed behaviour of Zw 229-15 cannot be attributed to observational issues and has physical significance.

\subsection[Discussion]{Discussion}\label{sec:Compare}

The optical variability properties Zw 229-15 has been extensively studied by multiple groups with various analysis techniques. \citet{Mushotzky11} discovered the inconsistency of light curve of Zw 229-15 with the damped random walk (DRW) of \citet{Kelly09} by analysing the PSD of the first $4$ quarters of the light curve. They reported values for the short time-scale log-PSD slope between $-2.96$ (Q$8$) and $-3.31$ (Q$6$) by fitting the estimates of the PSD directly. Similarly high values were reported by \citet{Carini12} who fit two PSD models to quarters $4$ through $7$ of the re-processing of light curve analysed here and found log-PSD slopes $-2.88$ and $-2.83$ for a knee and broken power-law model respectively.

The full \Kepler light curve of Zw 229-15 was analysed using PSD and C-ARMA analysis methods by \citet{Edelson14}. Unlike the re-processing of \citet{CariniWilliamsAAS} used by us and in \citet{Kasliwal15b}, the analysis in \citet{Edelson14} used a much larger pixel-aperture ($32$-pixels) to minimize spacecraft-induced variability due to thermally-induced focus variations and differential velocity aberration---known issues with \Kepler light curves \citep{Kinemuchi12}. At the same time, using such large masks results in crowding problems---faint background sources bleed flux into the aperture because of the large \Kepler PSF contaminating the light curve with their variability. The simple PSD analysis of the re-processed light curve suggests that on time-scales longer than $\sim 5.6$~d, the PSD is well modelled by a power law with log-slope $-2$. On shorter time-scales, the log-PSD slope is much higher ($-4.51$). There is an extra PSD component with log-PSD slope $-1.28$ that contributes mostly at the $1$~d time-scale. The C-ARMA analysis uses the same software package (\textsc{carma\_pack}) used by us and found a bend time-scale of $\sim 4$~d with log-PSD slopes $-1.99$ and $-3.65$ on longer and shorter time-scales. \citet{Edelson14} tested the light curve for moire pattern noise and found evidence to suggest that moire pattern noise is a significant contaminant in both the flagged and un-flagged quarters.

\citet{Kasliwal15} analysed several \Kepler MAST light curves that had been de-trended of spacecraft-induced variability by the \Kepler SOC pipeline using co-trending basis vectors to quantify light curve features common to targets across the \Kepler FOV. Using a structure function method to fit a bent power-law PSD model, \citet{Kasliwal15} found log-PSD slope $-2.7$ with characteristic de-correlation time-scale $27.5$~d.

Most recently, \citet{CariniWilliamsAAS} applied a PSD analysis to the full light curve studied here and found log-PSD slope $2.80 \pm 0.43$ with turnover time-scale $66.94^{+14.3}_{-11.8}$~d. The same log-PSD slope was found by \citet{Kasliwal15b} using the structure function approach of \citet{Kasliwal15} suggesting that the slope and time-scale found by \citet{CariniWilliamsAAS} is significant and not sensitive to the analysis technique.

Regardless of the mechanism for variability, certain time-scales of interest may be associated with a thin disk. The shortest time-scale is the light crossing time-scale
\begin{equation}\label{eq:tlc}
t_{\mathrm{lc}} \sim \frac{r_{\mathrm{lc}}}{c} < 11~\text{d},
\end{equation}
where $r_{\mathrm{lc}}$ is the maximum radius that can be ascribed to $t_{\mathrm{lc}}$ \citep{Peterson}, i.e., this time-scale places a (very conservative) upper limit on the size of the region that variations could originate from. The shortest time-scale on which variations should occur is the dynamical time-scale given by
\begin{equation}\label{eq:tdyn}
t_{\mathrm{dyn}} \sim \frac{r}{v_{\mathrm{Kep}}} = \left( \frac{r^{3}}{GM_{\mathrm{BH}}} \right)^{\frac{1}{2}} \sim 1.5 - 1500~\text{d},
\end{equation}
where $v_{\mathrm{Kep}}$ is the Keplerian rotation velocity. Perturbations to hydrostatic equilibrium in the vertical ($z$) direction are smoothed out on time-scale
\begin{equation}\label{eq:tz}
t_{\mathrm{z}} \sim \frac{H}{c_{S}} \sim t_{\mathrm{dyn}},
\end{equation}
which may be observable in the form of quasi-periodic variations in the light curve. Deviations from thermal equilibrium, caused by fluctuations in the local dissipation rate, are damped out on the time-scale
\begin{equation}\label{eq:ttherm}
t_{\mathrm{therm}} \sim \frac{c_{S}^{2}r^{2}}{v_{\mathrm{Kep}}^{2} \nu} \sim \frac{t_{\mathrm{dyn}}}{\alpha}  \sim 2~\text{d} - 5~\text{yr}.
\end{equation}
Under the effect of viscous torques, matter diffuses through the disc on the viscous time-scale,
\begin{equation}\label{eq:tvisc}
t_{\mathrm{visc}} \sim \frac{r^{2}}{\nu} \sim \frac{t_{\mathrm{dyn}}}{\alpha (H/r)^{2}} \sim 1 - 10~\text{yr}.
\end{equation}
The \citet{Barth11} reverberation mapping study of Zw 229-15 suggests that $M_{\mathrm{BH}} = 1.0^{+0.19}_{-0.24} \times 10^{7} M_{\sun}$ with $r_{\mathrm{Sch}} = 2GM/c^{2} \sim 3 \times 10^{10}$~m or $0.2$~au. The bolometric luminosity was found to be $L_{\mathrm{bol}} = 6.4 \times 10^{43}$~erg~s$^{-1}$ corresponding to $L / L_{\mathrm{Edd}} \sim 0.05$. Assuming disk radii of $100 r_{\mathrm{Sch}}$ ($20$~au) to $10000 r_{\mathrm{{Sch}}}$ ($2000$~au), Shakura Sunyaev $\alpha$ parameter value between $0.1$ and $0.5$, and disc height to radius ratios between $0.01$ and $0.1$, we can compute the various time-scales of equations~\eqref{eq:tlc} through~\eqref{eq:tvisc}.

The smallest viscous time-scale is just under $1$~yr, suggesting that viscous fluctuations generated by the $\alpha$ parameter variations of \citet{Lyubarskii97} are unlikely to be responsible for the observed variability. $\alpha$ parameter variations in the \citet{Lyubarskii97} theory result in inward propagating fluctuations characterized by correlations between variability in different bands with hard lags, i.e., shorter wavelength bands lag longer wavelength bands. This phenomenon is typically observed only in the X-rays \citep{Vaughan04,McHardy04b,Arevalo06}, whereas in the UVOIR the situation is reversed with longer wavelength bands lagging shorter wavelength bands \citep{Wanders97,Sergeev05}.

Estimates of the thermal time-scales can range between $\sim 3$~d ($r = 10^{2} r_{\mathrm{Sch}}$ \& $\alpha = 0.5$) and $\sim 5$~yr ($r = 10^{4} r_{\mathrm{Sch}}$ \& $\alpha = 0.1$), making it plausible to associate the $t_{\mathrm{e-fold}} = 69$~d time-scale with the thermal time-scale of the disc. This may imply that the variability observed by \Kepler in the optical is driven by the unobserved X-ray variability of Zw 229-15 \citep{Krolik91}. However, it is now known that while X-ray variability drives small scale fluctuations in optical variability on short time-scales, longer time-scale large amplitude fluctuations exist in the optical that are too energetic to be driven by re-processed X-ray emission \citep{Uttley03,Arevalo09}. There is strong evidence to suggest that there are \textit{two} sources of variability in the optical--reprocessing of X-ray emissions that drives short time-scale low-amplitude variability, and a process local to the optical emitting region of the disk that drives larger amplitude, longer time-scale changes \citep{Gaskell08}. While the \Kepler light curve of Zw 229-15 is long enough to probe the days--weeks time-scales associated with the re-processing of the X-ray variability, it may not be sensitive to the longer time-scale variability intrinsic to the optical.

The dynamical time-scale of the disk ranges between $\sim 1.6$ and $\sim 1600$~d making it plausible to associate the observed $t_{\mathrm{e-fold}}$ with this quantity as well. The dynamical time-scale characterizes states that are perturbed from dynamical equilibrium such as by global g-modes \citep{ReynoldsMiller09a,ReynoldsMiller09b}. The light crossing time-scale is smaller than $\sim 11$~d making it unsuitable to associate with any of the time-scales that we have measured.

We have successfully recovered the $\sim 5.6$~d time-scale found by \citet{Edelson14}. We interpret it as the time lag at which the Green's function peaks, i.e., the duration between the initial impulse driving a flux perturbation and the peak in emitted flux. We have found that there is some indication of excess power at very high frequencies ($> 1$~d$^{-1}$) though the observation noise level makes the detection dubious. We treat this excess power as belonging purely to the disturbance process that generates the stochastic variability, in which case it may be caused by the thermal noise of emitting medium. We identify the $67$~d time-scale reported by \citet{CariniWilliamsAAS} with the e-folding time-scale $t_{\mathrm{e-fold}}$ of a flux perturbation in the accretion disk of the AGN, i.e., it is the duration over which flux perturbations decay by a factor of $\mathrm{e}$. Finally, we point out that the exact slope of the log-PSD is a function of frequency. From the C-ARMA($2$,$1$) PSD in equation~\eqref{eq:CARMA21PSD}, we see that over a considerable range of frequencies, one may expect the log-PSD to have slope close to $-2$. At very high frequencies, the log-PSD approaches $-4$ as the $\nu^{4}$ term in the denominator of equation~\eqref{eq:CARMA21PSD} begins to dominate. At very low frequencies, we expect the PSD to flatten. This is exactly the behavior inferred by all the studies of Zw 229-15 and is predicted of the C-ARMA($2$,$1$) process. The reason why different studies have fit a range of slopes at different frequencies may be attributable to variations between the exact processing used but is likely primarily due to sensitivity to different parts of the PSD.



\section[Conclusion]{Conclusions}\label{sec:Conclusions}

We demonstrate that the Continuous-time AutoRegressive Moving Average (C-ARMA) model of \citet{Kelly14} results from linearization of arbitrary phenomenological non-linear differential equations for the flux emitted by the AGN accretion disk. Small perturbations of the total flux emitted by an AGN can be modelled in the linear-regime as a linear differential equation driven by noise or a C-ARMA process. Such processes consist of an $n^{\mathrm{th}}$-order linear differential equation for the flux perturbation (LHS) stochastically driven by a linear combination of Wiener increments (RHS). The driving impulses may characterize complex MHD processes such as MRI turbulence. We show how insights into the variability-driving physics can be obtained by examining the PSD of the flux impulses. We propose that the homogenized linear differential equation corresponding to the C-ARMA process governs how flux perturbations evolve after they have been generated. We show how the evolution of the flux perturbations can be characterized by the Green's function of the linear-differential equation.

We propose the use of a new representation of the C-ARMA process in state-space form based on the observable companion form of a linear time-invariant system. This representation puts the dynamics of the flux variations into the state-equation and possesses a very simple form for the observation equation. We argue that this representation is ideally suited for analyzing well-sampled flux light curves with constant sampling rate, especially in the presence of missing observations. Using the representation that we propose, the Kalman filtering and RTS smoothing equations can be applied to infer the values of the parameters the C-ARMA model and provide an estimate of the true light curve sans observation error.

We present a detailed analysis of the C-ARMA($2$,$1$) model or damped harmonic oscillator driven by violet noise. The C-ARMA($2$,$1$) process may be well suited for modelling AGN light curves based on light curves from the \Kepler mission and the Sloan Digital Sky Survey. We demonstrate how the Green's function increases steeply over a characteristic time-scale $t_{\mathrm{max}}$ before decaying exponentially with e-folding time-scale $t_{\mathrm{e-fold}}$. The driving disturbances of the C-ARMA($2$,$1$) model possess a flat PSD on long time-scales. On time-scales shorter than $t_{\mathrm{turn}}$, the PSD of the driving disturbances increases as $\nu^{2}$. We suggest that this short time-scale behaviour of the PSD may arise from the thermal motion of the material in the accretion disk.

We analyze a custom re-processing of the optical light curve of the Seyfert 1 galaxy Zw 229-15 presented in \citet{CariniWilliamsAAS} using the C-ARMA formalism. We find that the C-ARMA($2$,$1$) model of section~\ref{sec:CARMA21} well characterizes the observed flux variations. Fluctuations in the light curve of Zw 229-15 are strongly damped with damping ratio $\zeta = 3.3$. Thus, perturbations to the flux result in a smooth decay to the steady-state flux level with no oscillatory behavior. The Green's function of the flux perturbations is observed to peak on time-scale $t_{\mathrm{max}} = 5.6$~d which is consistent with the PSD time-scale reported by \citet{Edelson14}. After the initial peak, flux perturbations decay with e-folding time-scale $t_{\mathrm{e-fold}} = 69$~d that is consistent with the PSD time-scale reported by \citet{CariniWilliamsAAS}. We find that on time-scales shorter than $\sim 1$~d, the PSD of the disturbances rises steeply as $\nu^{2}$ as reported by \citet{Edelson14}, however the measurement noise level of \Kepler makes this finding tentative. We hunt for non-physical spacecraft-induced contributions to the observed behavior by examining the effect of moire-pattern noise on our results. While the moire-pattern noise affected quarters have slightly higher measurement noise than the clean quarters, the overall model-fit does not change significantly when excluding the noisy quarters. We conclude that the variability behavior observed in Zw 229-15 by \Kepler is consistent with an over-damped harmonic oscillator driven by a colored-noise process. We demonstrate how breaks and features found in conventional PSD analysis of light curves may be interpreted in the context of characteristic time-scales that arise from the C-ARMA($2$,$1$) model.

C-ARMA analysis techniques offer tremendous promise for studying AGN \& BHXRB variability. The C-ARMA formalism can be extended to model different accretion states using formalisms such the Threshold C-ARMA processes of \citet{DimensionEstimationBrockwell}. C-ARMA processes can also be extended to non-simultaneous multi-wavelength observations to probe the connection between the variability observed in different bands. Numerical simulations of accretion disks are already able to produce time-domain synthetic spectra \citep{Schnittman13b} and will soon robustly include radiative transfer in the modelling \citep{FragileAccretion}. C-ARMA models may be useful in comparing the synthetic spectra and light curves produced by numerical models of the accretion disk and winds to observational results. Moving forward, more sophisticated techniques such as the particle filter may be able to directly infer the underlying non-linear stochastic differential equations governing variability without the need for linearization \citep{Hanif15}. Tools for performing C-ARMA analysis are provided in the \textsc{c++} and \textsc{Python} package \href{https://github.com/AstroVPK/kali}{\textsc{k\={a}l\={i}}} and can be obtained from \url{https://github.com/AstroVPK/libcarma}.

\section*{Acknowledgements}

We acknowledge support from NASA grant NNX14AL56G. This paper includes data collected by the Kepler mission. Funding for the Kepler mission is provided by the NASA Science Mission directorate. Some of the data presented in this paper were obtained from the Mikulski Archive for Space Telescopes (MAST). STScI is operated by the Association of Universities for Research in Astronomy, Inc., under NASA contract NAS5-26555. Support for MAST for non-HST data is provided by the NASA Office of Space Science via grant NNX09AF08G and by other grants and contracts. \href{https://github.com/AstroVPK/kali}{\textsc{k\={a}l\={i}}} uses the optimization libraries provided by Steven G. Johnson, The NLopt nonlinear-optimization package, http://ab-initio.mit.edu/nlopt. We wish to thank Jack O'Brien, Jackie Moreno, Nate B. Lust and Adam Lidz for their input and discussions.





\bibliographystyle{mnras}
\bibliography{allrefs} 




\appendix

\section[Perturbation Evolution]{How Do Flux Perturbations Evolve?}\label{sec:LHS}

We assume that the variations in AGN luminosity may be ascribed to flux perturbations intrinsic to the accretion disk, corona, and winds. It is likely that variability originates from multiple mechanisms simultaneously. These mechanisms are almost certainly non-linear \citep{UMV05}. Hence the total flux emitted by the AGN evolves via a (probably non-linear) differential equation such as the integral of the surface mass density equation \citep[see][eq. 4]{LightmanEardley74} that governs how perturbations evolve. We shall examine how linear perturbations of the total flux evolve. Let the total flux emitted by an AGN be $F(t)$. Suppose $F$ obeys the $p^{\mathrm{th}}$-order non-linear differential equation
\begin{equation}\label{eq:FluxDE}
\frac{\mathrm{d}^{p}F}{\mathrm{d}t^{p}} = g(F,\frac{\mathrm{d}F}{\mathrm{d}t},\ldots,\frac{\mathrm{d}^{p-1}F}{\mathrm{d}t^{p-1}},\mathcal{F},t),
\end{equation}
where $\mathcal{F}$ is the contribution to the total flux from the appearance of hot- and cold-spots in the accretion disk. We shall attempt to linearise this system by probing the behaviour of a small perturbation in flux about a solution to this equation \citep{Wiberg,Stengel}. We begin by re-writing this n$^\mathrm{th}$-order non-linear differential equation for $F(t)$ as a series of coupled $1^\mathrm{st}$-order non-linear differential equations for $F(t)$ and its derivatives. Define
\begin{equation}\label{eq:GeneralRelationship}
F_{k}(t) = \frac{\mathrm{d}^{k-1}F(t)}{\mathrm{d}t^{k-1}},
\end{equation}
for $1 \leq k \leq n$. Since
\begin{equation}\label{eq:Linearize}
\frac{\mathrm{d}F_{k}}{\mathrm{d}t} = \frac{\mathrm{d}}{\mathrm{d}t}\left(\frac{\mathrm{d}^{k-1}F}{\mathrm{d}t^{k-1}}\right) = \frac{\mathrm{d}^{k}F}{\mathrm{d}t^{k}} = F_{k+1},
\end{equation} we have
\begin{equation}\label{eq:1stOrderSystem}
\begin{aligned}
\frac{\mathrm{d}F_{1}}{\mathrm{d}t} &= F_{2}, \\
&\ \ \vdots \\
\frac{\mathrm{d}F_{p-1}}{\mathrm{d}t} &= F_{p}.
\end{aligned}
\end{equation}
Then, we may rewrite equation~\eqref{eq:FluxDE} as
\begin{equation}\label{eq:1stOrderNonLinearDE}
\frac{\mathrm{d}F_{p}}{\mathrm{d}t} = g(F_{1},F_{2},\ldots,F_{p},\mathcal{F},t).
\end{equation}
Non-linear differential equations can be difficult to treat analytically. However, we are not interested in the behaviour of the total flux. Rather, we are interested in how small perturbations ($\delta F/F << 1$) from a known solution to the non-linear equations behave. In order to understand how small perturbations behave, we shall linearize this system of equations by looking at the form of the equations in the presence of a small perturbation. Starting from initial conditions $F_{1}(t_{0})$, $\ldots$, $F_{p}(t_{0})$ with input contributions $\mathcal{F}_{0}(t)$, suppose our system evolves as per the solutions $\phi_{1}(t)$, $\ldots$, $\phi_{p}(t)$, i.e., these solutions obey equation~\eqref{eq:1stOrderNonLinearDE} under the constraints of equation~\eqref{eq:1stOrderSystem}. $\phi_{1}(t)$ is the flux emitted by the object at time $t$ while $\phi_{k}$ with $1 < k \leq p$ are the $k^{\mathrm{th}}$ derivatives of the flux. Now consider the evolution of the system starting from the (slightly perturbed) initial conditions $F_{1}'(t_{0}) = F_{1}(t_{0}) + X_{1}(t_{0})$, $\ldots$, $F_{p}'(t_{0}) = F_{p}(t_{0}) + X_{p}(t_{0})$ where the various $X_{k}$ are small. Furthermore, we shall also subject the system to a slightly perturbed input  $\mathcal{F}_{0}'(t) =  \mathcal{F}_{0}(t) +  \mathcal{X}_{0}(t)$. Then, the perturbed solutions, $\phi_{1}(t) + \chi_{1}(t)$, $\ldots$, $\phi_{p}(t) + \chi_{p}(t)$, must continue to satisfy equation~\eqref{eq:1stOrderNonLinearDE} under the constraints of equation~\eqref{eq:1stOrderSystem}. We have introduced $\chi_{k}(t)$ with $1 \leq k \leq p$ as small perturbations to the original solutions $\phi_{k}$, i.e., $\chi_{1}(t)$ is a small perturbation to the flux $\phi_{1}(t)$ at time $t$ while the $\chi_{k}(t)$ are small perturbations of the derivatives of the flux. Starting with the constraint equations, we have
\begin{equation}\label{eq:1stOrderPerturnedSystem}
\begin{aligned}
\frac{\mathrm{d}(\phi_{1} + \chi_{1})}{\mathrm{d}t} = \frac{\mathrm{d}\phi_{1}}{\mathrm{d}t} + \frac{\mathrm{d}\chi_{1}}{\mathrm{d}t} &= \phi_{2} + \chi_{2},\\
&\ \ \vdots \\
\frac{\mathrm{d}(\phi_{p-1} + \chi_{p-1})}{\mathrm{d}t} = \frac{\mathrm{d}\phi_{p-1}}{\mathrm{d}t} + \frac{\mathrm{d}\chi_{p-1}}{\mathrm{d}t} &= \phi_{p} + \chi_{p},\\
\end{aligned}
\end{equation}
But since the $\phi_{k}$ satisfy equation~\eqref{eq:1stOrderSystem}, we have
\begin{equation}\label{eq:FinalSystem1}
\begin{aligned}
\frac{\mathrm{d}\chi_{1}}{\mathrm{d}t} &= \chi_{2}, \\
&\ \ \vdots \\
 \frac{\mathrm{d}\chi_{p-1}}{\mathrm{d}t} &= \chi_{p}.
\end{aligned}
\end{equation}
To linearize equation~\eqref{eq:1stOrderNonLinearDE}, we begin by noting
\begin{equation}\label{eq:TempStep01}
\frac{\mathrm{d}(\phi_{p} + \chi_{p})}{\mathrm{d}t} = \frac{\mathrm{d}\phi_{p}}{\mathrm{d}t} + \frac{\mathrm{d}\chi_{p}}{\mathrm{d}t} = g(F_{1}',F_{2}',\ldots,F_{p}',\mathcal{F}_{0}',t).
\end{equation}
We may Taylor expand the RHS of this equation about the $\phi_{1}$, ..., $\phi_{p}$ and $\mathcal{F}_{0}$ and cancel $\mathrm{d}\phi_{p}/\mathrm{d}t = f(\phi_{1},\phi_{2},\ldots,\phi_{p},\mathcal{F}_{0},t)$ to obtain
\begin{equation}\label{eq:FinalSystem2}
\frac{\mathrm{d}\chi_{p}}{\mathrm{d}t} = \left.{\frac{\partial g}{\partial F_{1}}}\right \vert_{\phi_{1}} \chi_{1} + \ldots + \left.{\frac{\partial g}{\partial F_{p}}}\right \vert_{\phi_{p}} \chi_{p} + \left.{\frac{\partial g}{\partial \mathcal{F}_{0}}}\right \vert_{\mathcal{F}_{0}}\mathcal{F}_{0}'.
\end{equation}
Equations~\eqref{eq:FinalSystem1} and \eqref{eq:FinalSystem2} are the linearised, $1^{\mathrm{st}}$-order system of equations that describe the evolution of a tiny perturbation of the flux about an exact solution. In general, the coefficients of the $\chi_{k}$ in equation~\eqref{eq:FinalSystem2} are functions of time and change as the system evolves. Let us consider the special case where the exact solution is the steady-state solution (if it exists). Under this restriction, equation~\eqref{eq:FinalSystem2} governing the evolution of a tiny perturbation in the flux simplifies to
\begin{equation}\label{eq:FinalSystem3}
\frac{\mathrm{d}\chi_{p}}{\mathrm{d}t} = a_{1} \chi_{1} + a_{2} \chi_{2} + \ldots + a_{n} \chi_{p} + b \mathcal{F}_{0}',
\end{equation}
where
\begin{equation}\label{eq:aCoefficients}
a_{k} = \left.{\frac{\partial g}{\partial F_{k}}}\right \vert_{\phi_{k}},
\end{equation}
and
\begin{equation}\label{eq:bCoefficients}
b = \left.{\frac{\partial g}{\partial \mathcal{F}_{0}}}\right \vert_{\mathcal{F}_{0}}.
\end{equation}
Therefore, we may re-write the linearised system of equations in equation~\eqref{eq:FinalSystem3} along with the constraints in equation~\eqref{eq:FinalSystem1} in the form of the n$^{\mathrm{th}}$-order linear differential equation
\begin{equation}\label{eq:FinalSystemLHS}
\frac{\mathrm{d}^{p}\chi}{\mathrm{d}t^{p}} + \alpha_{1} \frac{\mathrm{d}^{p-1}\chi}{\mathrm{d}t^{p-1}} + \ldots + \alpha_{p-1} \frac{\mathrm{d}\chi}{\mathrm{d}t} + \alpha_{p} \chi = u,
\end{equation}
where we have set $\alpha_{k} = -a_{k}$ and $u = b\mathcal{F}_{0}'$ for brevity. Using differentials, this equation may be written as
\begin{equation}\label{eq:FinalSystemLHSDiff}
\mathrm{d}^{p}\chi + \alpha_{1} \mathrm{d}^{p-1}\chi + \ldots + \alpha_{p-1} \mathrm{d}\chi + \alpha_{p} \chi = u,
\end{equation}
where $\mathrm{d}u$ is an increment of $u$. \textit{We see that the LHS of this equation is identical to the LHS of the C-ARMA process equation~\eqref{eq:CARMAIntro} suggesting that a C-ARMA process may be a good model for small flux variations.}

Dimensional consistency requires that
\begin{equation}\label{eq:ARUnits}
[\alpha_{k}] = T^{-k},
\end{equation}
where $T$ is time, i.e., the $\alpha_{k}$ have units that are powers of various frequencies. Since equation~\eqref{eq:FinalSystemLHS} governs how flux perturbations evolve over time, we shall refer to this equation as the `Dissipation Equation'. We define the characteristic polynomial of the LHS of this equation as
\begin{equation}\label{eq:ARCharPoly}
\alpha(z) = z^{p} + \alpha_{1} z^{p-1} + \ldots + \alpha_{p-1}z + \alpha_{p},
\end{equation}
with roots $\rho_{k}$. The flux perturbations are stable, i.e., do not increase without bound, if $\operatorname{Re}(\rho_{k}) < 0$. We shall posit a form for $u$ in section~\ref{sec:RHS}.

We find that if the total flux emitted by the accretion disk obeys a non-linear differential equation, we can linearize that equation to determine the behaviour of small perturbations of the flux from the steady-state solution. The linear differential equation that governs how a small flux perturbation evolves is identical to the C-ARMA process equation suggesting that such processes may be good models for the stochastic variability seen in AGN accretion disks. In the next section, we present a convenient physical interpretation of equation~\eqref{eq:FinalSystemLHS} using the method of the Green's function.

\section[Green's Function Computation]{How to Compute the Green's Function}\label{sec:GFComputation}

We compute the Green's function of equation~\eqref{eq:FinalSystemLHS} by finding a solution of the homogenized version of the equation. Consider the homogeneous differential equation corresponding to equation~\eqref{eq:FinalSystemLHS}
\begin{equation}\label{eq:HomoCARMA}
\frac{\mathrm{d}^{p}f}{\mathrm{d}t^{p}} + \alpha_{1} \frac{\mathrm{d}^{p-1}f}{\mathrm{d}t^{p-1}} + \ldots + \alpha_{p-1} \frac{\mathrm{d}f}{\mathrm{d}t} + \alpha_{p} f = 0.
\end{equation}
To obtain the Green's function solution of this equation, we shall drive this equation with a unit impulse, i.e., a Dirac delta function, located at $t = 0$
\begin{equation}\label{eq:GreenCARMA}
\frac{\mathrm{d}^{p}G}{\mathrm{d}t^{p}} + \alpha_{1} \frac{\mathrm{d}^{p-1}G}{\mathrm{d}t^{p-1}} + \ldots + \alpha_{p-1} \frac{\mathrm{d}G}{\mathrm{d}t} + \alpha_{p} G = \delta(t),
\end{equation}
where $G(t)$ is the desired Green's function solution. To find this solution, we shall first solve equation~\eqref{eq:HomoCARMA} by finding the roots of the characteristic polynomial of equation~\eqref{eq:ARCharPoly} and then apply boundary conditions based on the properties of $\delta(x)$. From the definition of the characteristic polynomial in equation~\eqref{eq:ARCharPoly}, the (distinct) roots are $\rho_{k}$ with $\operatorname{Re}(\rho_{k}) < 0$ for stability. Then the solution of the homogeneous equation~\eqref{eq:HomoCARMA} is
\begin{equation}\label{eq:HomoSol}
f(t) = \sum_{k = 1}^{n} c_{k} \mathrm{e}^{\rho_{k}t},
\end{equation}
where the $c_{k}$ are constants. Hence the solution of the inhomogeneous equation~\eqref{eq:GreenCARMA} may be found by using the continuity properties of $G(t)$ and its derivatives. For $t < 0$, $\mathrm{d}^{p-k}G/\mathrm{d}t^{p-k} = 0$ for $1 \leq k < p$ while at $t = 0$, $\mathrm{d}^{p}G/\mathrm{d}t^{p}$ has the same type of discontinuity as $\delta(t)$. Therefore $\mathrm{d}^{p-1}G/\mathrm{d}t^{p-1} = 1$ when $t = 0$, i.e., it behaves like the step function. Next, we have that $\mathrm{d}^{p-k}G/\mathrm{d}t^{p-k} = 0$ at $t = 0$ for all $1 < k \leq p$ since each successive lower-order derivative is a $(k-1)^{\mathrm{th}}$-order polynomial located at the origin. Following this line of reasoning, we arrive at the following boundary conditions
\begin{equation}\label{eq:BC1a}
 \left.\frac{\mathrm{d}^{p-1}G}{\mathrm{d}t^{p-1}}\right|_{t = 0} = 1,\\
\end{equation}
and
\begin{equation}\label{eq:BC2a}
 \left.\frac{\mathrm{d}^{p-k}G}{\mathrm{d}t^{p-k}}\right|_{t = 0} = 0,
\end{equation}
for all $2 \leq k \leq p$. Computation of the derivatives in equations~\eqref{eq:BC1a} and \eqref{eq:BC2a} results in
\begin{equation}\label{eq:BC1b}
 \sum_{k = 1}^{p} c_{k} \rho_{k}^{p-1} = 1,
 \end{equation}
 and
\begin{equation}\label{eq:BC2b}
 \sum_{k = 1}^{p} c_{k} \rho_{k}^{k-1} = 0,
\end{equation}
for all $2 \leq k \leq p$. The $c_{k}$ may then be found by solving the matrix equation
\begin{equation}\label{eq:BCs}
\left( \begin{array}{ccccc}
\rho_{1}^{p-1} & \rho_{2}^{p-1} & \hdots & \rho_{p-1}^{p-1} & \rho_{p}^{p-1} \\
\rho_{1}^{p-2} & \rho_{2}^{p-2} & \hdots & \rho_{p-1}^{p-2} & \rho_{p}^{p-2} \\
\vdots & \vdots & \ddots & \vdots & \vdots \\
\rho_{1} & \rho_{2} & \hdots & \rho_{p-1} & \rho_{p} \\
1 & 1 & \hdots & 1 & 1 \\
\end{array}\right) \left( \begin{array}{c} c_{1} \\ c_{2} \\ \vdots \\ c_{p-1} \\ c_{p} \end{array} \right) = \left( \begin{array}{c} 1 \\ 0 \\ \vdots \\ 0 \\ 0 \end{array} \right).
\end{equation}
In short, the Green's function for equation~\eqref{eq:FinalSystemLHS} is given by
\begin{equation}\label{eq:GreenSol}
G(t) = \sum_{k = 1}^{p} c_{k} \mathrm{e}^{\rho_{k}t},
\end{equation}
where the $c_{k}$ satisfy equation~\eqref{eq:BCs}.

\section[Stochastic Impulses Mathematics]{The Mathematics of the Driving Stochastic Impulses}\label{sec:RHS}

Fundamental to the treatment of the input stochastic perturbations, i.e., the $u$ of equation~\eqref{eq:FinalSystemLHSDiff} as some form of noise process is the concept of the Wiener process $W$, also known as Brownian motion \citep{Doob,Davis,Jacobs,Oksendal}. Wiener processes and processes derived from it are of importance in a number of areas of physics, engineering and finance \citep{Jacobs}. For example, the random motion of the electrons in a strip of metal produces a small, stochastically fluctuating potential known as Johnson noise \citep{Gillespie96} that can be modelled using Wiener increments. The Wiener process cannot be used directly as the driving process because the likelihood of a given value of the Wiener process is time-dependent, i.e., Wiener processes are not \textit{stationary}. The power spectral density (PSD) of the Wiener process decreases as the square of the frequency and so has excess power on long time-scales. Processes with a flat PSD are described in the engineering literature as `white' noise because they have equal power at all frequencies in analogy with white light possessing equal amounts of light of all colors. Analogously, processes with non-flat PSD are described as `colored' with various adjectives such as `blue', `red', and `pink'. being used to describe the relative amplitude of the PSD at various frequencies. In this sense, the Weiner processes is known in the engineering literature as `red'-noise. A better choice for the driving process is some linear combination of the differentials of $W$, i.e., increments of $dW$. We use differentials because mathematically, due to the discontinuous nature of $W$, the derivatives of $W$ are undefined. Writing $u$ as a linear combination of the differentials of $W$, we obtain
\begin{equation}\label{eq:FinalSystemRHSDiff}
u = \beta_{0} (\mathrm{d}W) + \beta_{1} \mathrm{d}(\mathrm{d}W) + \ldots + \beta_{p-2} \mathrm{d}^{p-2}(\mathrm{d}W) + \beta_{p-1} \mathrm{d}^{p-1}(\mathrm{d}W)
\end{equation}
where $\mathrm{d}W$ is the differential of the Wiener process and is mathematically well-defined. By definition
\begin{equation}\label{eq:WienerIncr}
W(t) = \int_{0}^{t} \mathrm{d}W.
\end{equation}
Dimensional consistency requires that
\begin{equation}\label{eq:MAUnits}
[\beta_{k}] = [\chi]~T^{k+\frac{1}{2}-p},
\end{equation}
where $\chi$ is in units of flux, e.g. $\mathrm{erg}~\mathrm{cm}^{-2}$. Analogous to equation~\eqref{eq:ARCharPoly}, we define the the characteristic polynomial of equation~\eqref{eq:FinalSystemRHSDiff} to be
\begin{equation}\label{eq:MACharPoly}
\beta(z) = \beta_{0} + \beta_{1} z + \ldots + \beta_{p-2} z^{p-2} + \beta_{p-1} z^{p-1},
\end{equation}
with roots $\mu_{k}$. For equation~\eqref{eq:FinalSystemRHSDiff} to be invertible, we require that $\operatorname{Re}(\mu_{k}) < 0$.

The power spectral density (PSD), $S_{uu}(\nu)$, of the process $u(t)$ is defined to be
\begin{equation}\label{eq:PSD}
S_{uu}(\nu) = \langle |\widetilde{u}(\nu)|^{2} \rangle,
\end{equation}
where
\begin{equation}\label{eq:FT}
\widetilde{u}(\nu) = \frac{1}{2\upi} \int_{-\infty}^{\infty}u(t)\mathrm{e}^{-2\upi \imath \nu t}\mathrm{d}t
\end{equation}
is the Fourier transform of $u(t)$, i.e., $\widehat{u}(\nu)$ denotes the Fourier transform of $u(t)$. The PSD of the Wiener increments is constant at all frequencies \citep{HandbookOfStatistics19Brockwell}, i.e.,
\begin{equation}\label{eq:WienerIncrementFT}
S_{\mathrm{d}W\mathrm{d}W} = \frac{1}{2 \upi},
\end{equation}
corresponds to infinite total power when integrated over all frequencies. Real processes only resemble the Wiener process and Wiener increments over an interval of frequencies known as the bandpass of the process. Outside this bandpass, the PSD is set to $0$. The PSD of the driving process in equation~\eqref{eq:FinalSystemRHSDiff} may be computed using
\begin{equation}\label{eq:FTDerivative}
\widetilde{\frac{\mathrm{d}^{k}u}{\mathrm{d}t^{k}}} = (2 \upi \imath \nu)^{k}\widetilde{u},
\end{equation}
in conjunction with equation~\eqref{eq:WienerIncrementFT} yielding
\begin{equation}\label{eq:RHSPSD}
S_{uu}(\nu) = \frac{1}{2 \upi } |\beta(2 \upi \imath \nu)|^{2},
\end{equation}
where $\beta$ is the moving average polynomial. The PSD of the driving impulses of equation~\eqref{eq:RHSPSD} is a polynomial in $\nu$. This implies that at high frequencies, the power contributed by this process is unbounded. However, consider what happens to the PSD when $\mathrm{d}W$ is integrated. The integral of the Wiener increments, i.e., the Wiener process itself, has PSD $S_{WW} \propto 1/\nu^{2}$. Successive integrations decrease the logarithmic slope of the PSD at high frequencies by a factor of $2$ each time. If $u$ is supplied as input to equation~\eqref{eq:FinalSystemLHS}, it is integrated $n$-times. If the overall process $\chi$ is to have finite power at high frequencies, we \textit{must} limit the highest-order of the derivative of $\mathrm{d}W$ in equation~\eqref{eq:FinalSystemRHSDiff} to $p-1$ so that after $p$-integrations, the overall process has PSD with negative logarithmic slope corresponding to finite high-frequency power, i.e., the bandpass of the driving disturbance process rapidly decreases at high frequencies. At low frequencies, we expect the power in our system to drop to $0$ as we approach the time-scale on which AGN are active $\sim 10^{7}$~yr. On the more relevant time-scales that can be probed by us, the PSD should essentially flatten to a constant at low frequencies. Finally, we may formulate our system by combining equations~\eqref{eq:FinalSystemLHSDiff} and~\eqref{eq:FinalSystemRHSDiff} into the It\={o} differential equation
\begin{multline}\label{eq:CARMA}
\mathrm{d}^{p}\chi + \alpha_{1} \mathrm{d}^{p-1}\chi + \ldots + \alpha_{p-1} \mathrm{d}\chi + \alpha_{p} \chi = \\ \beta_{0} (\mathrm{d}W) + \beta_{1} \mathrm{d}(\mathrm{d}W) + \ldots + \beta_{p-1} \mathrm{d}^{p-1}(\mathrm{d}W),
\end{multline}
with $\operatorname{Re}(\rho_{k}) < 0$ and $\operatorname{Re}(\mu_{k}) < 0$ where the $\rho_{k}$ and $\mu_{k}$ are the roots of the autoregressive- and moving average- polynomials of equations~\eqref{eq:ARCharPoly} and~\eqref{eq:MACharPoly} respectively. It\={o} differential equations arise when the solution of the differential equation requires the integration of a stochastic function against Wiener increments. Such equations can be solved by using the methods of the It\={o} calculus \citep{Oksendal}. Given the form of the C-ARMA process, we can write the PSD of the C-ARMA process \citep{HandbookOfStatistics19Brockwell} as
\begin{equation}\label{eq:CARMAPSD}
S_{\chi\chi}(\nu) = \frac{1}{2 \upi }\frac{|\beta(2 \upi \imath \nu)|^{2}}{|\alpha(2 \upi \imath \nu)|^{2}}.
\end{equation}

In practice not all of the moving average terms, $\beta_{k}$, may be present. In particular, if the first $k$ moving average terms are absent, i.e., $\beta_{l} = 0$ for $l < k+1$, the RHS has $q = p - k$ terms. A stochastic processes of this form is known in the statistical inferencing community as a Continuous-time AutoRegressive Moving Average process of order $p$ and $q$ or C-ARMA($p$,$q$) process \citep{DimensionEstimationBrockwell,HandbookOfStatistics19Brockwell,Brockwell14,Kelly14}. The maximum allowed value of $q$ is dictated by the value of $p$, i.e., $q < p$.

The PSD of equation~\eqref{eq:CARMAPSD} can be expressed as
\begin{equation}\label{eq:CARMAPSDForm1}
S_{\chi\chi}(\nu) = \frac{1}{2 \upi }\frac{\sum_{i,j = 0}^{p} (2 \upi \imath \nu)^{2p - i -j} \beta_{i} \beta_{j} (-1)^{p - j}}{\sum_{m,n = 0}^{q} (2 \upi \imath \nu)^{2q - m -n} \alpha_{m} \alpha_{n} (-1)^{q - n}},
\end{equation}
with $\alpha_{0} \equiv 1$. Consider the terms with $2p - i - j$ odd, i.e., complex terms. Since $2p$ is always even, either $i$ or $j$ must be odd. While $(2 \upi \imath \nu)^{2p - i -j}$ and $\beta_{i} \beta_{j}$ are symmetric in $i$ \& $j$, $(-1)^{p - j}$ is not. Interchanging $i$ \& $j$ leads us to the conclusion that odd (complex-valued) terms occur in pairs $(2 \upi \imath \nu)^{2p - i -j}\beta_{i} \beta_{j}$ and $-(2 \upi \imath \nu)^{2p - j -i}\beta_{j} \beta_{i}$ that cancel each other. We see that the numerator is purely real and consists of a sum of terms that are proportional to even powers of frequency. The same logic applies to the denominator.

Defining $k = 2q - i - j$ \& $l = 2p - m - n$ allows us to rewrite equation~\eqref{eq:CARMAPSDForm1} as
\begin{multline}\label{eq:CARMAPSDForm1}
S_{\chi\chi}(\nu) = \frac{N(\nu)}{D(\nu)} = \\ \frac{1}{2 \upi } \frac{\sum_{i = 0}^{q}\sum_{k = 2q - i}^{3q - i} (2\upi\nu)^{k} \imath^{k} (-1)^{3q - i - k} \beta_{i}\beta_{k + i - 2q}}{\sum_{m = 0}^{p}\sum_{l = 2p - m}^{3p - m} (2\upi\nu)^{l} \imath^{l} (-1)^{3p - m - l} \alpha_{m}\alpha_{l + m - 2p}},
\end{multline}
from which we see that the $\nu^{k\mathrm{-th}}$ power term in the numerator is given by
\begin{equation}\label{eq:PSDNumerator}
N_{\nu^{k}}(\nu) = \sum_{i = 0}^{q} (2\upi\nu)^{k} \imath^{k} (-1)^{3q - i - k} \beta_{i}\beta_{k + i - 2q},
\end{equation}
and the $\nu^{l\mathrm{-th}}$ power term in the denominator is given by
\begin{equation}\label{eq:PSDDenominator}
D_{\nu^{l}}(\nu) = \sum_{m = 0}^{p} (2\upi\nu)^{l} \imath^{l} (-1)^{3p - m - l} \alpha_{m}\alpha_{l + m - 2p}.
\end{equation}

The Wiener-Khintchine Theorem lets us compute the corresponding \textit{Auto-Covariance Function} (ACVF) of the observed light curve \citep{HandbookOfStatistics19Brockwell} as
\begin{equation}\label{eq:CARMAACVF}
\gamma_{\chi\chi}(\tau) = \int_{-\infty}^{\infty}\chi(t)\chi(t + \tau) dt = \sum_{k = 1}^{p} \frac{\beta(\rho_{k}) \beta(-\rho_{k})}{\alpha(\rho_{k})\alpha(-\rho_{k})}\mathrm{e}^{\rho_{k} \tau},
\end{equation}
where the $\rho_{k}$ are the roots of the autoregressive characteristic polynomial.

\section[\href{https://github.com/AstroVPK/kali}{\textsc{k\={a}l\={i}}}]{How Does \href{https://github.com/AstroVPK/kali}{\textsc{k\={a}l\={i}}} Infer C-ARMA Parameters}\label{sec:kali}

\subsection[Fitting C-ARMA Processes]{How to Fit a C-ARMA Process to Observations?}\label{sec:Fitting}

We wish to infer the values of the co-efficients in equation~\eqref{eq:CARMAIntro} that defines a particular C-ARMA process in order to compute the physically interesting Green's function and driving disturbance PSD for the observed light curve. While many techniques can be applied to perform this task, we use the Kalman filter \citep{Kalman60, Simon} since it requires trivial matrix inversions and determinant calculations ($1 \times 1$ matrices in the case of single-filter observations). An added advantage of the Kalman filter is that it is designed to allow us to update the likelihood calculation when new measurements become available. Finally, the framework in which the Kalman filter operates, the state-space framework, is also easy to extend to the case of multi-band, non-simultaneous observations of AGN as will become available with next generation surveys such as PanSTARRS and the LSST.

The Kalman filter allows us to obtain Bayesian posteriors for the distribution function of the C-ARMA coefficients. These coefficients can then be used to compute physically meaningful time-scales that will yield insight into accretion physics. Inferencing linear systems using the Kalman filter has been successfully applied to control complex systems including the Apollo program for over $50$ years \citep{GrewalAndrews10}, and has since then come into vogue in fields as diverse as biostatistics \citep{Jones} and econometrics \citep{DurbinKoopman,Harvey}.

Application of the Kalman filter requires that the system be represented in \textit{state-space} form. The central concept underlying the application of the Kalman filter is that of the state of the system--an abstract quantity that completely characterizes the system at a given instant e.g. the position and velocity of a rocket. This state evolves via the state-equation--a differential (continuous-time) or difference (discrete-time) equation that tells us how to update the state of the system from time-step to time-step. Generically, the state-equation consists of three components--(1) a deterministic isolated time evolution component; (2) a deterministic input response component; and (3) a random disturbance component. The input component can be used to \textit{control} the system by driving it to a desired state. One of the key benefits of the Kalman approach is that the state is not observed directly--instead the system is monitored via the observation equation e.g. we observe the angular position of the rocket. The act of observation is necessarily imprecise and hence the observed quantities are contaminated with noise. The Kalman filter is used to recover estimates of the state of the system given the noisy observations of the state and the form of the state- and observation-equations.


We represent the C-ARMA process in continuous-time state-space form \citep{Wiberg,Friedland} as
\begin{equation}\label{eq:StateEqn}
\mathrm{d}\mathbfit{x} = \mathbfss{A}\mathbfit{x}\mathrm{d}t + \mathbfit{B}\mathrm{d}w,
\end{equation}
and
\begin{equation}\label{eq:ObsEqn}
\chi(t) = \mathbfss{H(t)}\mathbfit{x} + v,
\end{equation}
where $\mathbfit{x}$ is the system-state vector, $W$ is the Wiener process of section~\ref{sec:RHS}, $v \sim \mathcal{N}(0,\sigma^{2}_{N})$ is the measurement noise, and $\chi$ is the (noisy) observed flux at time $t$. The matrix $\mathbfss{A}$ is known as the `state-transition matrix', while the matrix $\mathbfss{H}$ is known as the `observation matrix'. There are infinitely many ways of writing a C-ARMA process in state-space form--we shall discuss the two most relevant \citep{Denham74,Kailath74} because of their use in common packages for C-ARMA inferencing such as \href{https://github.com/AstroVPK/kali}{\textsc{k\={a}l\={i}}} (this work) and \textsc{carma\_pack} \citep{Kelly14}. All other forms may be obtained from these `canonical' forms via similarity transforms.

The `$1^{\mathrm{st}}$ canonical form' of \citet{Wiberg} which is also the `$2^{\mathrm{nd}}$ companion form' of \citet{Friedland}, puts the dynamics of the C-ARMA process solely into the state-equation by representing equation~\eqref{eq:CARMA} using
\begin{multline}\label{eq:1stCanForm}
\mathbfss{A} = \left( \begin{array}{ccccc}
-\alpha_{1} & 1 & \hdots & 0 & 0 \\
-\alpha_{2} & 0 & \hdots & 0 & 0 \\
\vdots & \vdots & \ddots & \vdots & \vdots \\
-\alpha_{p-1} & 0 & \hdots & 0 & 1 \\
-\alpha_{p} & 0 & \hdots & 0 & 0 \\
\end{array}\right); \\ \mathbfit{B} = \left( \begin{array}{c} \beta_{p-1} \\ \beta_{p-2} \\ \vdots \\ \beta_{1} \\ \beta_{0} \end{array} \right); \mathbfss{H} = \left( \begin{array}{c} 1 \\ 0 \\ \vdots \\ 0 \\ 0 \end{array} \right)^{\top}.
\end{multline}

The `$2^{\mathrm{nd}}$ canonical form' or `phase-variable canonical form of \citet{Wiberg} which is the `$1^{\mathrm{st}}$ companion form' of \citet{Friedland} (see also \citealp{Kelly14} and references therein), breaks-up the dynamics of the C-ARMA process across the state- and observation-equations by representing equation~\eqref{eq:CARMA} using
\begin{multline}\label{eq:2ndCanForm}
\mathbfss{A} = \left( \begin{array}{ccccc}
0 & 1 & 0 & \hdots & 0 \\
0 & 0 & 1 & \hdots & 0 \\
\vdots & \vdots & \vdots & \ddots & \vdots \\
0 & 0 & 0 & \hdots & 1 \\
-\alpha_{p} & -\alpha_{p-1} & -\alpha_{p-2} & \hdots & -\alpha_{1} \\
\end{array}\right); \\ \mathbfit{B} = \left( \begin{array}{c} 0 \\ 0 \\ \vdots \\ 0 \\ 1 \end{array} \right); \mathbfss{H} = \left( \begin{array}{c} \beta_{0} \\ \beta_{1} \\ \vdots \\ \beta_{p-2} \\ \beta_{p-1} \end{array} \right)^{\top}.
\end{multline}

Note that the components of the state-vector $\mathbfit{x}$ are not the same in the two representations but carry the same information, i.e., we are projecting the abstract state-vector into a particular coordinate system. Regardless of the choice of representations used, we can integrate the state-equation \citep{DimensionEstimationBrockwell,Oksendal} to obtain an update equation for $\mathbfit{x}$
\begin{equation}\label{eq:State}
\mathbfit{x}(t + \delta t) = \mathbfit{x}(t)\mathrm{e}^{\mathbfss{A}\delta t} + \int_{0}^{\delta t} \mathrm{e}^{\mathbfss{A}(\delta t-u)}\mathbfit{B} \mathrm{d}W,
\end{equation}
where $\int_{0}^{\delta t} \mathrm{e}^{\mathbfss{A}(\delta t-u)}\mathbfit{B} \mathrm{d}W$ is an It\={o} integral. The presence of the stochastic It\={o} integral implies that we cannot predict \textit{exactly} what the future values of $\mathbfit{x}$ will be because the integral in equation~\eqref{eq:State} contributes randomly for every realization of the process. However, we can predict the expectation value of the integral and the $1-\sigma$ confidence limit on that expectation value \citep[chapter 6]{Davis}. These are given by
\begin{equation}\label{eq:ExpVal}
\left\langle \int_{0}^{\delta t} \mathrm{e}^{\mathbfss{A}(\delta t-u)}\mathbfit{B} \mathrm{d}W \right\rangle = \mathbfit{0},
\end{equation}
and
\begin{multline}\label{eq:CovVal}
\left\langle \int_{0}^{\delta t} \mathrm{e}^{\mathbfss{A}(\delta t-u)}\mathbfit{B} \mathrm{d}W \int_{0}^{\delta t} \mathrm{e}^{\mathbfss{A}(\delta t-u)}\mathbfit{B} \mathrm{d}W \right\rangle = \\ \int_{0}^{\delta t} \mathrm{e}^{\mathbfss{A}\xi}\mathbfit{B}\mathbfit{B}^{\top}\mathrm{e}^{\mathbfss{A}^{\top}\xi} \mathrm{d}\xi,
\end{multline}
allowing us to rewrite equation~\ref{eq:State} as
\begin{equation}\label{eq:IntegratedState}
\mathbfit{x}(t + \delta t) = \mathbfss{F}\mathbfit{x}(t) + \mathbfit{w}_{k},
\end{equation}
with \begin{equation}\label{eq:TransMat}
\mathbfss{F} = \mathrm{e}^{\mathbfss{A}\delta t},
\end{equation}
and $\mathbfit{w}_{k} \sim \mathcal{N}(\mathbfit{0},\mathbfss{Q})$ with
\begin{equation}\label{eq:DistMat}
\mathbfss{Q} = \int_{0}^{\delta t} \mathrm{e}^{\mathbfss{A}\xi}\mathbfit{B}\mathbfit{B}^{\top}\mathrm{e}^{\mathbfss{A}^{\top}\xi} d\xi.
\end{equation}
$\mathbfss{F}$ is called the transition matrix and $\mathbfss{Q}$ is the disturbance variance-covariance matrix.

Computing the matrices $\mathbfss{F}$ and $\mathbfss{Q}$ is readily accomplished using the eigen-decomposition of $\mathbfss{A}$. Let
\begin{equation}
\mathbfss{A} = \mathbfss{v}\mathbfss{w}\mathbfss{v}^{-1},
\end{equation}
where $\mathbfss{v}$ is the matrix whose $i^{\mathrm{th}}$ column is the eigenvector $\mathbfit{v}_{1}$ of $\mathbfss{A}$ and $\mathbfss{w}$ is the diagonal matrix whose diagonal elements are the corresponding eigenvalues, i.e., $\mathbfss{w}_{i,i} = w_{i}$. If the eigenvalues are distinct, then $\mathbfss{F}$ may be computed as
\begin{equation}\label{eq:F}
\mathbfss{F} = \mathrm{e}^{\mathbfss{A}\delta t} = \mathbfss{v}\mathrm{e}^{\mathbfss{w}\delta t}\mathbfss{v}^{-1},
\end{equation}
where we exploit the property that the matrix exponential of a diagonal matrix is simply the matrix of the exponentials of the diagonal elements, i.e.,
\begin{equation}\label{eq:MatrixExp}
\mathrm{e}^{\mathbfss{w}\delta t} = \left( \begin{array}{ccccc}
\mathrm{e}^{w_{1} \delta t} & 0 & \hdots & 0 & 0 \\
0 & \mathrm{e}^{w_{2} \delta t} & \hdots & 0 & 0 \\
\vdots & \vdots & \ddots & \vdots & \vdots \\
0 & 0 & \hdots & \mathrm{e}^{w_{p-1} \delta t} & 0 \\
0 & 0 & \hdots & 0 & \mathrm{e}^{w_{p} \delta t} \\
\end{array}\right).
\end{equation}
To compute $\mathbfss{Q}$, we rewrite equation~\ref{eq:DistMat} substituting in the expression for the matrix exponential of $\mathbfss{A}$ from equation~\ref{eq:F}.
\begin{multline}\label{eq:DistMat2}
\mathbfss{Q} = \int_{0}^{\delta t} \mathrm{e}^{\mathbfss{A}\xi}\mathbfit{B}\mathbfit{B}^{\top}\mathrm{e}^{\mathbfss{A}^{\top}\xi} d\xi = \\ \mathbfss{v} \int_{0}^{\delta t} \mathrm{e}^{\mathbfss{w}\xi}\mathbfss{v}^{-1}\mathbfit{B}\mathbfit{B}^{\top}(\mathbfss{v}^{-1})^{\top} \mathrm{e}^{\mathbfss{w}\xi} d\xi \mathbfss{v}^{\top},
\end{multline}
where we have factored the matrices $\mathbfss{v}$ and $\mathbfss{v}^{\top}$ out of the integral. Defining $\mathbfss{C} = \mathbfss{v}^{-1}\mathbfit{B}\mathbfit{B}^{\top}(\mathbfss{v}^{-1})^{\top}$ lets us rewrite the integral as
\begin{equation}\label{eq:DistMat3}
\mathbfss{Q} = \mathbfss{v} \int_{0}^{\delta t} \mathrm{e}^{\mathbfss{w}\xi}\mathbfss{C}\mathrm{e}^{\mathbfss{w}\xi} d\xi \mathbfss{v}^{\top}.
\end{equation}
Let the integral be $\mathbfss{D}$ with $i^{\mathrm{th}}$ \& $j^{\mathrm{th}}$ entries
\begin{equation}\label{eq:DistMat4}
\mathbfss{D}_{i,j} = \int_{0}^{\delta t} \mathrm{e}^{(w_{i} + w_{j})\xi} \mathbfss{C}_{i,j} d\xi = \frac{\mathrm{e}^{(w_{i} + w_{j})\delta t} - 1}{w_{i} + w_{j}}\mathbfss{C}_{i,j} .
\end{equation}
Then we calculate
\begin{equation}\label{eq:DistMat5}
\mathbfss{Q} = \mathbfss{v}\mathbfss{D}\mathbfss{v}^{\top},
\end{equation}

We also define
\begin{equation}\label{eq:DistMat6}
\mathbfss{J}_{i,j} = \int_{0}^{\infty} \mathrm{e}^{(w_{i} + w_{j})\xi} \mathbfss{C}_{i,j} d\xi = \frac{ - 1}{w_{i} + w_{j}}\mathbfss{C}_{i,j}.
\end{equation}
and use it to compute
\begin{equation}\label{eq:Sigma}
\boldsymbol{\Sigma} = \mathbfss{v}\mathbfss{J}\mathbfss{v}^{\top},
\end{equation}
which can be used to compute the ACVF of the process via \citep{DimensionEstimationBrockwell}
\begin{equation}\label{eq:ACVFStateSpace}
\gamma_{\chi\chi}(T) = \mathbfss{H} \mathbfss{v}\mathrm{e}^{\mathbfss{w}T}\mathbfss{v}^{-1} \boldsymbol{\Sigma}\mathbfss{H}^{\top}.
\end{equation}

These equations for the computation of $\mathbfss{F}$, $\mathbfss{Q}$, and $\boldsymbol{\Sigma}$ allow us to use the Kalman filter to perform C-ARMA simulations \& perform inferencing.

\subsection[Representations]{Equivalent Representations of C-ARMA Processes}

We have introduced C-ARMA processes using equation~\eqref{eq:CARMAIntro}. A particular C-ARMA process can be specified via the co-efficients of the LHS and RHS of equation~\eqref{eq:CARMAIntro}. That is to say that if $\Theta$ is the set of C-ARMA process parameters, then $\Theta$ can be written
\begin{equation}\label{eq:Theta}
\Theta = \{ \alpha_{1}, \alpha_{2}, \ldots, \alpha_{p-1}, \alpha_{p}, \beta_{0}, \beta_{1}, \ldots, \beta_{q-1}, \beta_{q} \}.
\end{equation}
A second, equivalent, representation can be obtained by factorizing the C-AR and C-MA polynomials of equation~\eqref{eq:ARCharPoly} and equation~\eqref{eq:MACharPoly}. However, the C-MA polynomial is scaled by the co-efficient of the highest-order term $\beta_{q}$, suggesting that we define the \textit{modified} C-MA polynomial to be
\begin{equation}\label{eq:modMACharPoly}
\beta'(z) = \frac{\beta_{0}}{\beta_{q}} + \frac{\beta_{1}}{\beta_{q}} z + \ldots + \frac{\beta_{q-1}}{\beta_{q}} z^{q-1} + z^{q},
\end{equation}
with roots $\mu_{k}$. The corresponding component of the state-space representation is
\begin{equation}\label{eq:BPrime}
\mathbfit{B}' = \left( \begin{array}{cccccccc} 0 & \hdots & 0 & 1 & \frac{\beta_{q-1}}{\beta_{q}} & \hdots & \frac{\beta_{1}}{\beta_{q}} & \frac{\beta_{0}}{\beta_{q}} \end{array} \right)^{\top}.
\end{equation}
Define $\mathbfss{C}' = \mathbfss{v}^{-1}\mathbfit{B}'\mathbfit{B}'^{\top}(\mathbfss{v}^{-1})^{\top}$ and $\mathbfss{J}'_{i,j} = \frac{ - 1}{w_{i} + w_{j}}\mathbfss{C}'_{i,j}$ to compute $\boldsymbol{\Sigma}' = \mathbfss{v}\mathbfss{J}'\mathbfss{v}^{\top}$. Clearly, $\mathbfss{C} = \beta_{q}^{2}\mathbfss{C}'$, leading to the useful expression
\begin{equation}\label{eq:SigmaSigmaPrime}
\boldsymbol{\Sigma} = \beta_{q}^{2}\boldsymbol{\Sigma}',
\end{equation}
where $\boldsymbol{\Sigma}[0,0]$ is the true variance of the C-ARMA process and $\boldsymbol{\Sigma}'[0,0]$ is the variance of the C-ARMA process with unit $\beta_{q}$. $\boldsymbol{\Sigma}'[0,0]$ can be computed from the C-AR roots $\rho_{k}$ and the C-MA roots $\mu_{k}$ with no knowledge of $\beta_{q}$ required. If we specify $\boldsymbol{\Sigma}[0,0]$, we can recover $\beta_{q}$ via equation~\eqref{eq:SigmaSigmaPrime}. Therefore, an equivalent representation, $\Rho$ of the C-ARMA process is
\begin{equation}\label{eq:Rho}
\Rho = \{ \rho_{1}, \rho_{2}, \ldots, \rho_{p-1}, \rho_{p}, \mu_{1}, \ldots, \mu_{q-1}, \mu_{q}, \boldsymbol{\Sigma}[0,0] \}.
\end{equation}

\subsection[Sampling]{Sampling Patterns}\label{sec:Sampling}

Observation cadences can be broadly classified into two groups - those with regular sampling (RS) and those with irregular sampling (IRS). RS patterns occur when observations are made at fixed intervals to within the precision of the clock governing the schedule. We can relate the number of observations $N$ made to the duration of the sampling $T$ in terms of the sampling interval $\delta t = T/N$. The times at which the process is sampled $t$ must then be given by $t = n \delta t$ for $1 \leq n \leq N$ and can therefore be indexed by $n$. Occasionally, natural or mission-induced events make it impossible to observe the object of interest for one or more simultaneous cadences. Such incidents are called `missing-observations'. We track missing-observations via a mask $\mathcal{M}_{n}$. If the $n^{\mathrm{th}}$ observation is missing, $\mathcal{M}_{n} = 0$. Otherwise, $\mathcal{M}_{n} = 1$. Suppose the light curve under consideration is regularly observed (e.g. with a space-based telescope such as Kepler) with missing-observations, then we shall write the integrated state-equation~\eqref{eq:IntegratedState} as
\begin{equation}\label{eq:ARMAStateEq}
\mathbfit{x}_{k+1} = \mathbfss{F}\mathbfit{x}_{k} + \mathbfit{w}_{k},
\end{equation}
where $\mathbfit{w}_{k} \sim \mathcal{N}(\mathbfit{0},\mathbfss{Q})$. The original observation equation~\eqref{eq:ObsEqn} remains unchanged through this process but is now discrete
\begin{equation}\label{eq:ARMAObsEqn}
\chi_{n} = \mathbfss{H}_{n}\mathbfit{x}_{n} + v_{n},
\end{equation}
where $v_{n} \sim \mathcal{N}(0,\sigma_{N,n}^{2})$ are (heteroskedastic) measurement errors. We recognize equations~\eqref{eq:ARMAStateEq} and~\eqref{eq:ARMAObsEqn} as a discrete-time state-space system and identify it with the \textit{Auto-Regressive Moving Average} (ARMA) processes of \citet{BrockwellDavisTSTM}, \citet{BrockwellDavisITSF}, \citet{DurbinKoopman}, and in astronomical variability studies of \citet{Scargle81,GaskellPeterson87,KoratkarGaskell91}. If the light curve has missing observations, we recommend that the $1^{\mathrm{st}}$ canonical form from equation~\eqref{eq:1stCanForm} be chosen. The benefit of this form of the state-space equations is that the observation matrix is very simple. To correctly handle missing-observations, it suffices to set $\mathbfss{H}_{n} = \left( \begin{array}{ccccc} \mathcal{M}_{n} & 0 & \hdots & 0 & 0 \end{array} \right)$ and the measurement uncertainty of the $n^{\mathrm{th}}$ observation $\sigma_{N,n}^{2} = \infty$ \citep{Jones80,Jones}.

If the light curve is obtained with an IRS pattern as is usually the case with ground-based astronomical observations such as those carried out by the Sloan Digital Sky Survey (SDSS), we define $\delta t_{n} = t_{n+1} - t_{n}$. If the IRS pattern is fairly dense, it may be advantageous to simply determine an artificial fixed sampling interval $\delta t$ as the greatest common divisor of the individual $\delta t_{n}$. Then we add missing-observations into the light curve so that the missing-observations correspond to the gaps in the original light curve. At this point, we may treat the formerly IRS light curve as a RS light curve with lots of missing observations.

If the IRS pattern is fairly sparse, it is computationally cheaper to recompute the transition matrix $\mathbfss{F}_{n}$ and the disturbance variance-covariance matrix $\mathbfss{Q}_{n}$ at every time step as proposed by \cite{JonesAckerson90}, \cite{Jones} and \cite{Kelly14}. The exact IRS pattern for which the two methods are equally fast depends on the machine architecture and hence our implementation of the C-ARMA modelling algorithm in \href{https://github.com/AstroVPK/kali}{\textsc{k\={a}l\={i}}} allows for both methods to be used at different time-steps within the same light curve for optimal speed, i.e., long gaps or sections can be treated as unevenly sampled data while occasional gaps are treated as missing values.

\subsection[Parameter Estimation]{Parameter Estimation with the Kalman Filter}\label{sec:Kalman}

Having obtained the discrete-time state- and observation-equations for the original C-ARMA process, we may now use the Kalman filter to compute the likelihood of having observed the light curve given the C-ARMA model. This likelihood can then be used to determine the likelihood of the C-ARMA model given the data using Bayes' rule. The future goal will then become to determine a suitable informative prior using a combination of previous observations and physical insight.

As stated earlier, one of the many attractive properties of the Kalman filter is that it operates sequentially on the light curve measurements--the Kalman filter uses the state-equation to \textit{predict} the current state of the system based on the previous state before the observation is made available to the algorithm, i.e., it computes an estimate of the \textit{a priori} state of the system $\widehat{\mathbfit{x}}^{-}_{n}$, where the `$^{-}$' signifies that it is an a priori estimate. Once the $n^{\mathrm{th}}$-observation is made available to the algorithm, the Kalman filter uses the observation to \textit{correct} the \textit{a priori} state of the system and thereby estimates the \textit{a posteriori} state of the system $\widehat{\mathbfit{x}}^{+}_{i}$, where the `$^{+}$' signifies that it is an a posteriori estimate. We can use the estimated a priori states of the system $\widehat{\mathbfit{x}}^{-}_{n}$ to predict estimates of the observed flux $\widehat{y}_{n} = \mathbfss{H}\widehat{\mathbfit{x}}^{-}_{n}$. The difference between the a priori predictions of the observed flux and the actual values of the observed flux, $n_{n}$, are known as `innovations'
\begin{equation}\label{eq:InnovationPrelim}
r_{n} = y_{n} - \widehat{y}_{n},
\end{equation}
and may be shown to be independent and identically distributed normal deviates \citep[chapter 10.2]{Simon} that are distributed according to $\mathcal{N}(0,S_{n})$ where $S_{n}$ is the estimate of the variance of the $n^{\mathrm{th}}$-innovation.

Suppose we wish to calculate the log-likelihood of the C-ARMA($p$,$q$) model $\mathcal{C}$ for the set of model parameters $\alpha_{k}$ and $\beta_{k}$ given the observed light curve of an object, $\{y_{n}\}$. From Bayes` theorem,
\begin{equation}\label{eq:ModelGivenData}
\ln \mathcal{L}(\mathcal{C}|\{y_{n}\}) = \ln \mathcal{L}(\{y_{n}\}|\mathcal{C}) + \ln \mathcal{L}(\mathcal{C}).
\end{equation}
The log-likelihood of the observed light curve $\{y_{n}\}$ given the model $\mathcal{C}$ may be computed using the innovations via
\begin{equation}\label{eq:DataGivenModel}
\ln \mathcal{L}(\{y_{n}\}|\mathcal{C}) = \sum_{n = 1}^{N} -\frac{\mathcal{M}_{n}}{2}(\ln 2 \upi + \ln S_{n} + \frac{r_{n}^{2}}{S_{n}}),
\end{equation}
where we include the value of the mask $\mathcal{M}_{n}$ to ensure that missing-observations do not contribute to the log-likelihood. For the prior, i.e., the likelihood of the model itself, we use the piecewise function
\begin{equation}\label{eq:Prior}
\ln \mathcal{L}(\mathcal{C}) =
  \begin{cases}
      \hfill 0    \hfill & \text{if~} \operatorname{Re}(\rho_{k}) < 0 \text{~and~} \operatorname{Re}(\mu_{k}) < 0 \text{~for all~} k  \\
      \hfill -\infty \hfill & \text{otherwise}
  \end{cases},
\end{equation}
where $\rho_{k}$ and $\mu_{k}$ are the roots of the autoregressive and moving average polynomials. This prior, while uninformative for acceptable values of the roots, ensures that the C-ARMA process corresponding to the roots is both stable and invertible.

The innovations required for the calculation of the light curve likelihood in equation~\eqref{eq:DataGivenModel} may be computed by applying the Kalman filter to the observed flux measurements. To do so, we require an estimate of the initial state of the system $\widehat{\mathbfit{x}}^{+}_{0}$ as well as the uncertainty in the initial state of the system $\mathbfss{P}^{+}_{0}$. The best estimate of the state of the system when initializing the Kalman filter is
\begin{equation}\label{eq:x0}
\widehat{\mathbfit{x}}^{+}_{0} = \mathbfit{0}
\end{equation}
and the best estimate of the state uncertainty is
\begin{equation}\label{eq:P0}
\mathbfss{P}^{+}_{0} = \boldsymbol{\Sigma},
\end{equation}
where $\boldsymbol{\Sigma}$ is computed using equations~\eqref{eq:DistMat6} and \eqref{eq:Sigma}.

Then starting with the first observation, i.e., $n = 1$, we iterate through the $N$ observations using the Kalman filter. We begin by computing the expected value of the a priori state:
\begin{equation}\label{eq:APrioriState}
\widehat{\mathbfit{x}}^{-}_{n} = \mathbfss{F} \widehat{\mathbfit{x}}^{+}_{n-1}.
\end{equation}
and the expected value of the a priori state uncertainty:
\begin{equation}\label{eq:APrioriUncertainty}
\mathbfss{P}^{-}_{n} = \mathbfss{F} \mathbfss{P}^{+}_{n-1} \mathbfss{F}^{\top} + \mathbfss{Q}.
\end{equation}
At this point, the innovation for this observation and the uncertainty in the innovation can be computed by comparing the expected observation to the actual observation using
\begin{equation}\label{eq:Innovation}
r_{n} = y_{n} - \mathbfss{H}\widehat{\mathbfit{x}}^{-}_{n},
\end{equation}
and
\begin{equation}\label{eq:InnovationVar}
S_{n} = ( \mathbfss{H}_{n} \mathbfss{P}^{-}_{n} \mathbfss{H}_{n}^{\top} + \sigma^{2}_{N,n}),
\end{equation}
where $\sigma^{2}_{N,n}$ is the (heteroskedastic) variance of the observation noise.

Before we may compute the innovation for the next observation, we must use our knowledge of the observed value of the AGN flux to update our knowledge of the state and the uncertainty in the state. Since the measurements are noisy, we must compute a gain to apply to the measurement. The gain should reflect the noisiness of the measurements, i.e., if the $n^{\mathrm{th}}$-observation is very noisy, we would like the gain that we apply to be very small. More significance is given to the observation as compared to the a priori state estimate if the gain is large. The optimal gain is the Kalman gain given by
\begin{equation}\label{eq:KalmanGain}
\mathbfss{K}_{n} = \mathbfss{P}^{-}_{n} \mathbfss{H}_{n}^{\top} S_{n}^{-1}.
\end{equation}
While any gain value will yield an unbiased estimate of the state, the Kalman gain yields an unbiased \textit{minimum variance} a posteriori estimate of the state and state uncertainty \citep[chapter 3.3]{Simon}. The Kalman gain is used to compute the a posteriori state and state covariance via
\begin{equation}\label{eq:APosterioriState}
\widehat{\mathbfit{x}}^{+}_{n} = \widehat{\mathbfit{x}}^{-}_{n} + \mathbfss{K}_{n} r_{n},
\end{equation}
and
\begin{equation}\label{eq:APosterioriUncertainty}
\mathbfss{P}^{+}_{n} = (\mathbfss{I} - \mathbfss{K}_{n} \mathbfss{H}_{n}) \mathbfss{P}^{-}_{n} (\mathbfss{I} - \mathbfss{K} _{n} \mathbfss{H}_{n})^{\top} + \mathbfss{K}_{n} \sigma^{2}_{n} \mathbfss{K}_{n}^{\top}.
\end{equation}
The innovations computed using this procedure may then be used along with equations~\eqref{eq:DataGivenModel} and~\eqref{eq:Prior} to compute the log-likelihood of the C-ARMA model $\mathcal{C}$ given the observed light curve $\{y_{n}\}$ using equation~\eqref{eq:ModelGivenData}. A multitude of numerical algorithms, including the Expectation-Maximization algorithm, can be used to maximize this likelihood. Alternatively, the full probability distribution of the C-ARMA model parameters can be probed by applying Markov Chain Monte Carlo (MCMC) techniques to the likelihood function.

We have discussed how the Kalman filter equations can be used to iterate through observations to compute the likelihood of having observed the light curve given a particular C-ARMA model. Bayes' rule allows us to use this likelihood to pick the most likely C-ARMA model. For now, we use a prior to select the portion of model space allowed by the rules of the C-ARMA process (the roots of the AR and MA polynomials must have negative real part). In the future, more informative priors can be used based on existing data, simulations and accretion theory.

\subsection[Model Selection]{How to Pick a Good C-ARMA Process?: Model Selection}\label{sec:ModelSelection}

Although we have reasons to expect AGN light curves to be well modelled by simple C-ARMA models of low order, it is prudent to search a large portion of the space of allowable models and use a statistically sound information criteria to pick the optimal model order. Information criteria such as the corrected Akaike Information Criteria (AICc) and the Deviance Information Criteria (DIC) balance model likelihood against model simplicity. The model that minimizes a chosen information criteria is the most-likely maximally-parsimonious model that fits the observed light curve \citep{ModelSelection}.

If a non-linear optimization algorithm is used to maximize the likelihood in equation~\eqref{eq:ModelGivenData}, we have point estimates of the model parameters and a single value for the likelihood of each model. To select the model order, we may use the AICc
\begin{equation}\label{eq:AICc}
AICc = 2(p+q+1) - 2 \ln \mathcal{L}(\mathcal{C}|\{y{n}\}) + \frac{2(p+q+1)(p+q+2)}{N-p-q-2},
\end{equation}
where $p$ and $q$ are the orders of the C-ARMA($p$,$q$) model and $N$ is the number of observations in the light curve.

Markov Chain Monte Carlo (MCMC) can be used to sample the likelihood function of equation~\eqref{eq:ModelGivenData}. Sampling the likelihood function yields reliable estimates of the confidence intervals of the C-ARMA model parameters even if the likelihood function is highly non-Gaussian. If multiple samples from the likelihood function are available, it is possible to use the DIC to select model order. The DIC is given by
\begin{equation}\label{eq:DIC}
DIC = \mu_{D(\mathcal{C})} + \frac{1}{2}\sigma^{2}_{D(\mathcal{C})},
\end{equation}
where the deviance $D(\mathcal{C})$ is computed as
\begin{equation}\label{eq:Deviance}
D(\mathcal{C}) = -2\ln \mathcal{L}(\mathcal{C}|\{y{n}\}).
\end{equation}
Either one of the DIC or the AICc may be used to select the optimum model order.

We have shown that regardless of the form of the equation governing the flux emitted by AGN and without a detailed analytic prescription for the form of accretion disk instabilities, it is possible to model small stochastic fluctuations in the observed light curve of AGN as a C-ARMA process. The physics of the accretion process may be probed by examining the Green's function and the Disturbance PSD of the inferred C-ARMA process. We have presented a method for inferring the model parameters of the C-ARMA process from observations of the light curve using the Kalman filter that is particularly suitable for light curves with regular sampling.

\subsection[Light Curve Smoothing]{What was the Light Curve Actually Doing?: Light Curve Smoothing}\label{sec:Smoothing}

After having selected a model and having determined the posterior probability distribution of the model parameters (or if using a point estimator, the maximum likelihood parameter estimates), the question arises: Can we estimate the true value of the light curve at any given instant regardless of the availability of an observation at the desired instant? This problem is known as the fixed-interval smoothing problem in the stochastic control and time series literature \citep{Simon,DurbinKoopman}. Optimal smoothing can be performed with the the Rauch-Tung-Striebel (RTS) smoother \citep*{RTS65}. The RTS smoother is based on the Kalman filter but requires additional steps after the standard Kalman filter has been applied. For astrophysical applications, smoothing can be important when comparing light curves from the same object obtained non-simultaneously in different filter bands. Smoothed light curves can be used to forecast the light curves of objects for short durations into the future. Such forecasts may be useful for survey optimization.

Assuming that the light curve has been regularly sampled (with missing observations), RTS smoothing is performed by first running the standard Kalman filter of section~\ref{sec:Kalman} in the forward direction (increasing time). In the application of the Kalman filter for computing the likelihood of the model, the a posteriori estimates of the state ($\widehat{\mathbfit{x}}^{+}_{n}$) and state covariance ($\mathbfss{P}^{+}_{n}$) for the $n^{\mathrm{th}}$-observation are not required once the a priori estimates of the state ($\widehat{\mathbfit{x}}^{-}_{n+1}$) and state-covariance ($\mathbfss{P}^{-}_{n+1}$) of the next state have been computed. To perform RTS smoothing, all the a priori and a posteriori state and state-covariance estimates must be retained for later use during the smoothing phase of the algorithm, i.e., to provide the best estimate of the state and state uncertainty, the algorithm uses \textit{all} the observations simultaneously instead of just the previous observations.
After the Kalman filter has been used to estimate the state and state-covariance for all the observations, we have the a priori and a posteriori estimates of the state and state covariance for all the (regularly-spaced) observations, i.e., $\{ \widehat{\mathbfit{x}}^{-}_{n} \}$, $\{ \widehat{\mathbfit{x}}^{+}_{n} \}$, $\{ \mathbfss{P}^{-}_{n} \}$, and $\{ \mathbfss{P}^{+}_{n} \}$ for $0 \leq n \leq N$.

The RTS smoothed estimates ($\widehat{\mathbfit{x}}_{n}$ and $\mathbfss{P}_{n}$) can be obtained by initializing the RTS smoother with
\begin{equation}\label{eq:xRTSInit}
\widehat{\mathbfit{x}}_{N} = \widehat{\mathbfit{x}}^{+}_{N},
\end{equation}
and
\begin{equation}\label{eq:PRTSInit}
\mathbfss{P}_{N} = \mathbfss{P}^{+}_{N}.
\end{equation}
Then iterating through the observations backwards (i.e., decreasing time), we first compute the Kalman gain
\begin{equation}\label{eq:KalmanRTS}
\mathbfss{K}_{n} = \mathbfss{P}^{+}_{n} \mathbfss{F}^{\top} (\mathbfss{P}^{-}_{n+1})^{-1},
\end{equation}
after which we can compute the smoothed state and state covariance using
\begin{equation}\label{eq:xRTS}
\widehat{\mathbfit{x}}_{n} = \widehat{\mathbfit{x}}^{+}_{n} + \mathbfss{K}_{n} (\widehat{\mathbfit{x}}_{n+1} - \widehat{\mathbfit{x}}^{-}_{n+1}),
\end{equation}
and
\begin{equation}\label{eq:PRTS}
\mathbfss{P}_{n} = \mathbfss{P}^{+}_{n} + \mathbfss{K}_{n} (\mathbfss{P}_{n+1} - \mathbfss{P}^{-}_{n+1}) \mathbfss{K}_{n}^{\top}.
\end{equation}
The smoothed state vector ($\widehat{\mathbfit{x}}_{n}$) can then be used to calculate the best estimate of the underlying light curve using the noiseless-observation equation
\begin{equation}\label{eq:NoiselessObsEq}
\chi_{n} = \mathbfss{H} \widehat{\mathbfit{x}}_{n},
\end{equation}
where we have used a fixed observation matrix that \textit{never} produces missing observations rather than the mask-dependent observation matrix $\mathbfss{H}_{n}$.

In practice, RTS smoothing is only performed after the best C-ARMA model and model parameters have been determined, i.e., it is not performed as part of the estimation but instead is a product after the estimation has been completed.





\bsp	
\label{lastpage}
\end{document}